\documentclass[aps,prl,twocolumn,superscriptaddress,showpacs,english]{revtex4-1}

\usepackage[T1]{fontenc}
\usepackage[latin9]{inputenc}
\usepackage{babel}
\usepackage{amsmath}
\usepackage{amssymb}
\usepackage{wasysym}
\usepackage{graphicx}
\usepackage{xcolor}
\usepackage{graphicx}

\usepackage[linktocpage=true,
  colorlinks=true, 
  pdfborder={0 0 0},
  linkcolor=blue,
  citecolor=red,
  filecolor=yellow,
  urlcolor=blue,
  bookmarks,
  pdfauthor={},
]{hyperref}

\newcommand{\Graz}{Institute of Theoretical and Computational Physics, Graz University of Technology, NAWI Graz, 8010 Graz, Austria}
\newcommand{\Oxford}{Department of Materials, University of Oxford, Parks Road, Oxford OX1 P3H, United Kingdom}
\newcommand{\hs}{H$_3$S}

\newcommand{\lis}{Li$_3$S}
\newcommand{\mus}{$\mu^*$}
\newcommand{\tc}{$T_C$}
\newcommand{\ef}{$E_F$}

\newcommand{\bcc}{$Im$\={3}$m$ }

\begin{document}

\title{Search for high-\tc \ conventional superconductivity at megabar pressures in the lithium-sulfur system}

\author{Christian Kokail}       \affiliation{\Graz}
\author{Christoph Heil}         \affiliation{\Oxford}
\author{Lilia Boeri}            \affiliation{\Graz}

\date{\today}

\begin{abstract}
{Motivated by the recent report of superconductivity above 200 K 
in ultra-dense hydrogen sulfide, we search for high-\tc\  
conventional superconductivity in the phase diagram 
of the binary Li-S system, using {\em ab-initio} methods for crystal structure
prediction and linear response calculations for the electron-phonon coupling.
We find that at pressures higher than 20 GPa, several new compositions,
besides the known Li$_2$S, are stabilized; many exhibit electride-like
{\em interstitial} charge localization observed in other 
alkali metal compounds. Of all predicted phases, only Li$_3$S 
at P > 640 GPa displays a sizable \tc, in contrast to what is observed
in sulfur and phosphorus hydrides, where several stoichiometries
lead to high \tc. We attribute this difference 
to 2$s$-2$p$ hybridization and avoided core overlap, 
and predict similar behavior for other alkali metal compounds.
} 
\end{abstract}

\pacs{~}
\maketitle

The successful prediction of a record critical temperature (\tc) of 203\,K in hydrogen sulfide (\hs) 
at 200\,GPa~\cite{DrozdovEremets_Nature2015,Troyan_Science_2016,Duan_SciRep2014} 
gave a considerable impulse to the {\em ab-initio} design of new
high-\tc\  superconductors at extreme pressures.
\hs \ was in fact the first example of a {\em conventional} 
high-temperature  superconductor whose crystal structure and \tc \
were first predicted completely from first-principles, and later 
confirmed experimentally.
It is now understood that its record-high \tc \ stems
from the constructive interference of large vibrational frequencies,
electronic van-Hove singularities at the Fermi level and 
large electron-phonon ($ep$) matrix
elements due the formation of {\em covalent} H-S
bonds.~\cite{Duan_SciRep2014,
SH_PRB-Mazin-2015,Heil-Boeri_PRB2015,FloresSanna_H3Se_EJPB2016,
quan_impact_2015,LiYanmingMa_JCP2014,Papaconstantopoulos_H3S_PRB2015,
Ortenzi_TB_2015,Errea_anhaPRL2015}
A few months after \hs,  a  high-\tc \ superconducting phase was
also found in compressed phosphines, 
which is compatible with several {\em metastable} PH$_x$ phases identified by first-principles calculations.~\cite{Drozdov_ph3_arxiv2015,shamp_decomposition_2015,Flores_PH3_PRBR2016,Fu_Ma_pnictogenH_2016} 
Recent reports of metallization in hydrogen at $\sim$ 350\,GPa have risen
the hope to attain superconductivity at room temperatures, or 
even higher.~\cite{Ashcroft_PRL1968,Cudazzo_PRL2008,mcmahon_high_2011,Borinaga_H_arxiv2016,
Szcz_superconducting_2009,Eremets_H_arxiv2016,Dalladay_H_Nature2016} 
While several hydrides have been proposed as prospective 
superconductors along the lines of \hs;~\cite{Ashcroft_PRL2004,tse_novel_2007,gao_high-pressure_2010,kim_general_2010,Yao_PNAS2010,Disilane_JAFL} 
 high-\tc\ superconductivity at high pressures in hydrogen-free
compounds is still a largely unexplored field.

In this work we search for high-\tc\  superconductivity at extreme pressures
in the  Li-S system, using the \textsc{USPEX} method for
{\em ab-initio} evolutionary crystal structure 
prediction,~\cite{oganovuniversal}
and density functional perturbation theory (DFPT)
 calculations of the $ep$ coupling
as implemented in \textsc{Quantum Espresso}.~\cite{QE_details,QE-2009} 
The underlying idea is to explore a hydrogen-free system similar to \hs;
Li-S is a natural choice, because lithium
belongs to the same group as hydrogen (similar chemical properties) 
and has a small atomic mass (large phonon frequencies).
At ambient pressure, Li-S is stable in crystalline form in the Li$_2$S composition; this compound has applications in lithium-based batteries, and has been investigated by several 
authors.~\cite{grzechnik2000reversible,Lazicki_LiS_PRB2006}
We find that at high pressures several new phases are stabilized,
many of which behave quite differently from the corresponding hydrides;
in particular, superconductivity is harder to attain, and the typical \tc's are much lower.
We will show that this can be explained by the different chemistry of the two elements, 
caused by the presence or absence of core electrons.~\cite{naumov2015chemical} 

In fact, lithium passes through a sequence of transitions under pressure 
from close-packed, metallic structures 
to open, semi-metallic or semiconducting ones.~\cite{hanfland2000new,shi2006theoretical,christensen2006calculated,marques_Lidense_PRL2011}
The increasing covalency is induced by the growing $2s$-$2p$ hybridization, and is
accompanied by the characteristic phenomenon of {\em interstitial charge localization},
i.e. the electronic valence charge tends to localize in interstitial regions of the
crystal to minimize the overlap with underlying 
atomic core states ({\em avoided core overlap}).~\cite{Ashcroft_interstitial_PRL2008,PhysRevLett.103.115501,PhysRevLett.104.216404,PhysRevLett.106.095502,Oganov_CaLi2_PRL2010}
Hydrogen, whose $1s$ valence electrons have no underlying core and  are well separated in energy 
from 2$p$ states, has a completely different behavior, transforming from molecular, insulating 
to close-packed metallic structures at very high pressures.~\cite{pickard_structure_2007} The highest \tc's in the two elements range from $\sim$ 16$\,$K, measured in lithium at 100\;GPa,~\cite{yao2009superconductivity,Profeta_LiKAl_PRL2006}
 to $\sim 350\,$K predicted for
hydrogen in the metallic phase.~\cite{Ashcroft_PRL1968,Cudazzo_PRL2008,mcmahon_high_2011,Borinaga_H_arxiv2016}

\begin{figure}[t]
\includegraphics[width=1.0\columnwidth,angle=0]{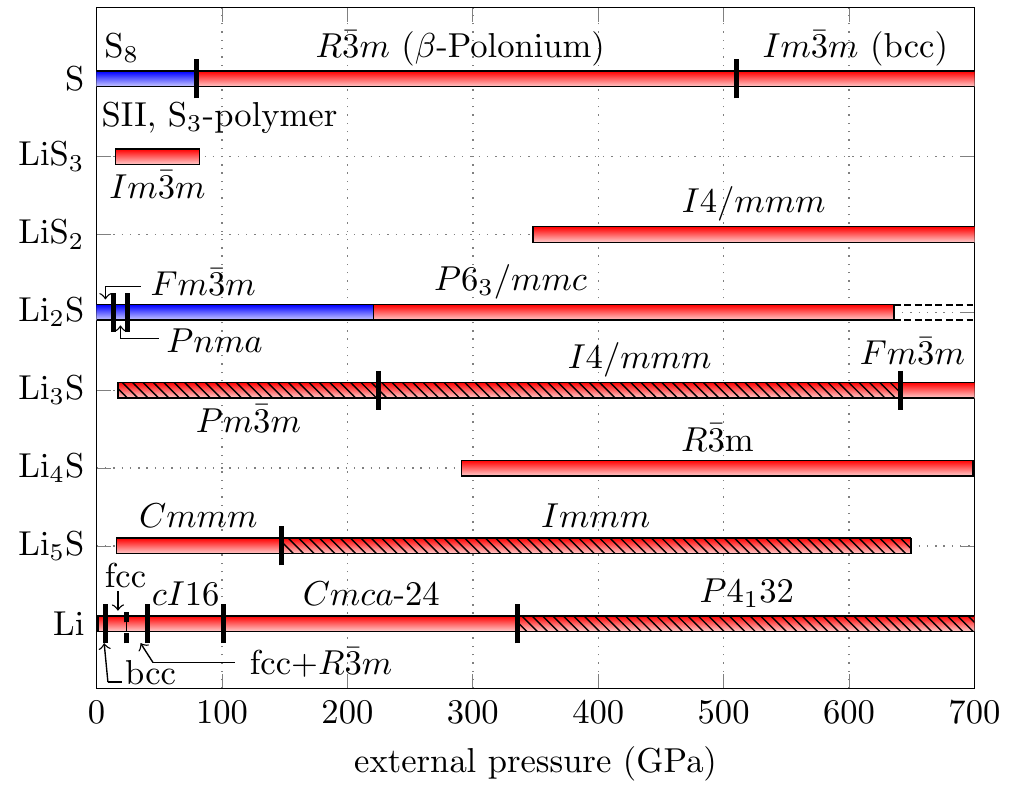}
\caption{(Color online)
Phase diagram for the Li-S system predicted by our DFT-EA
search. The vertical bars delimit regions of stability. Blue and red bars 
indicate insulating and metallic phases, respectively.
Shaded areas indicate phases with interstitial charge localization ({\em see text}).
}
\label{fig:PD}
\end{figure}

Figure~\ref{fig:PD} shows our theoretical phase diagram for the Li-S system,
constructed with a four-step procedure.
\mbox{(i) First}, we performed a preliminary scan of the phase space, with variable-compositions evolutionary algorithm (EA) runs from 0 to 
600$\,$GPa at 50$\,$GPa intervals.  (ii) For the most promising phases, \textit{i.e.} those which lie on the convex hull, we ran additional calculations at fixed compositions with 100$\,$GPa 
intervals starting from 0$\,$GPa to identify thermodynamically stable phases.
(iii)  The best three structures from each fixed composition run were relaxed once more with stricter convergence parameters, to ensure the correct enthalpy hierarchy of the phases. 
(iv) The best individuals  were relaxed further with a tighter convergence threshold, at pressure intervals of 50-100$\,$GPa, and 
 the resulting energy vs. volume curves were then fitted to a Murnaghan equation of state.~\cite{murnaghan1944compressibility}
This allowed us to obtain analytical expressions for the  enthalpy vs. pressure relation
for all structures, from which we could extract accurate stability ranges for all phases.~\cite{USPEX_details,PD_details}
Note that, in order to maintain our search within a reasonable time limit, 
we restricted the search space to phases with a maximum of 24 atoms 
per unit cell and 6 atoms per formula unit ($f.u.$).

\begin{figure}[t]
\includegraphics[width=1.0\columnwidth,angle=0]{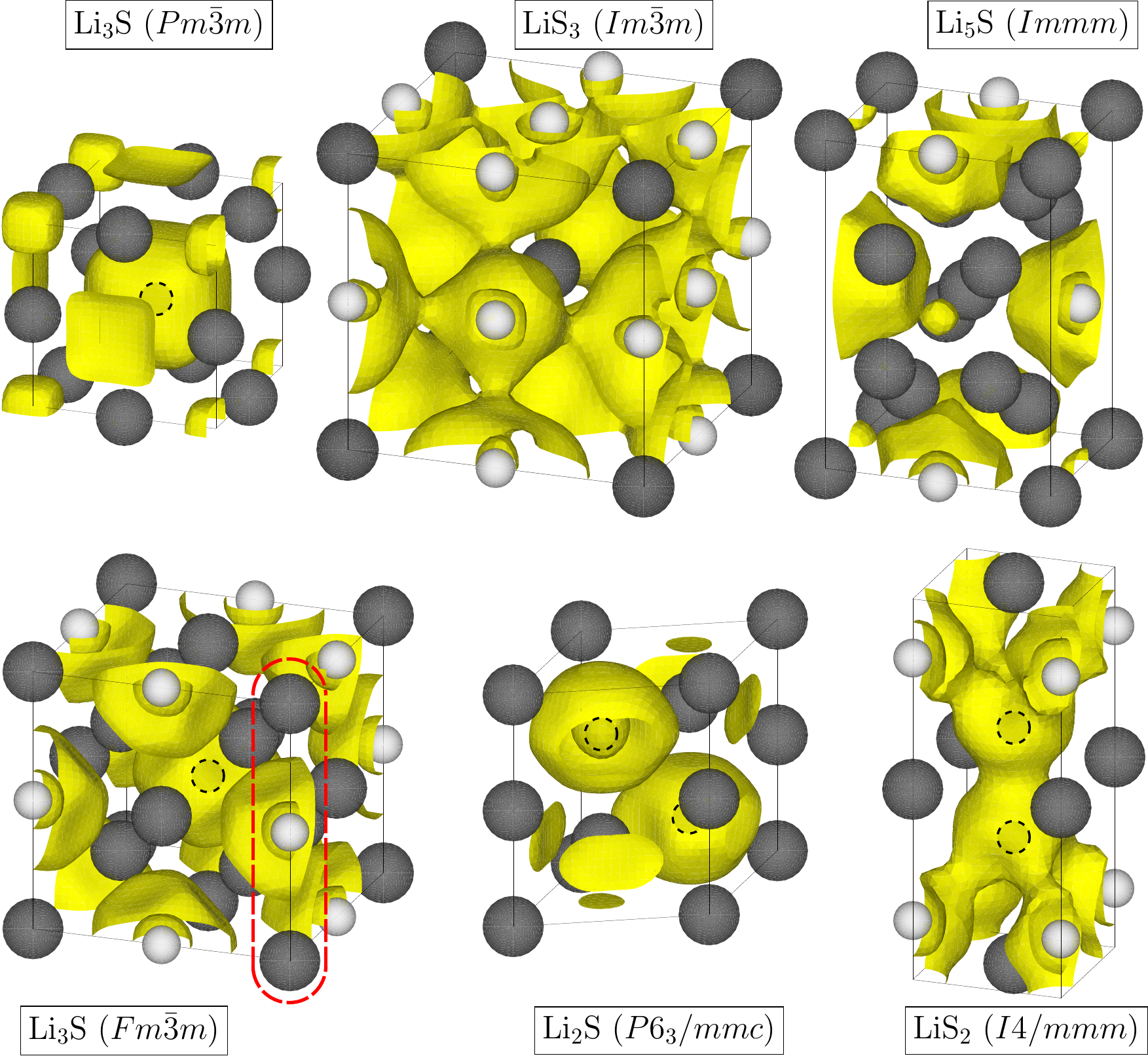}
\caption{(Color online): Crystal structure and isocontour (0.65) of the ELF for 
(top): Li$_3$S - $Pm\bar{3}m$ (100 GPa), LiS$_3$ - $Im\bar{3}m$ (100 GPa),  Li$_5$S - $Immm$ (500 GPa),  
(bottom) Li$_3$S - $Fm\bar{3}m$ (500 GPa), Li$_2$S - $P6_3/mmc$ (500 GPa)   and LiS$_2$ - $I4/mmm$ (500 GPa).
The structures are shown in scale; Li and S atoms are shown as black and white spheres, respectively.
Black, dashed circles indicate the location of S hidden by isosurfaces; red, dashed lines indicate important
bonds ({\em see text}).}
\label{fig:elfs}
\end{figure}

Before discussing the new phases found in our EA 
search, we note that our calculations reproduce
accurately literature results for the known phases, that is, for  
the two end members and Li$_2$S.
For elemental lithium, we find essentially the same phase diagram as Ref.~\onlinecite{ma2008high},
i.e. we predict a transition from a $bcc$ to a $fcc$ phase and then
 into a $cI16$ phase, 
stable until 100 GPa.~\cite{hanfland2000new} At 100 GPa, a $Cmca$ phase with 24 atoms
is stabilized, and remains the lowest in enthalpy up to  $\sim$330$\,$GPa, 
where a simple cubic (\textit{P}4$_1$32) phase occurs. 
Our results are in very good agreement with previous works which employ
unit cells up to 24 atoms,~\cite{shi2006theoretical,christensen2006calculated}
while recent calculations with larger supercells predicted two additional 
phases between 60 and 270 GPa -- \textit{Aba}2-40 (40 atoms per cell) and \textit{Cmca}-56 (56 atoms per cell). 
Having verified  that the enthalpy differences with respect to the \textit{Cmca}-24 phases are minimal
and do not affect the calculated convex hull, we decided to use the  \textit{Cmca}-24 phase in the whole range.
For sulfur, we predict a transition from the S$_8$ $\alpha$ phase to the polymeric S-II (2 GPa) and S$_3$-polymer (20 GPa)
phases.~\cite{oganov_sulphur}
At 80 GPa, the S$_3$-polymer phase transforms into the $\beta$-Po phase, which is metallic. 
At very large pressures ($\sim$510$\,$GPa) 
we predict a transition to a standard $bcc$ phase, as in Ref.~\onlinecite{zakharov1995theory}.
In agreement with previous calculations, and despite several attempts, we did not find any indication of a $bco$ phase as seen by experiments between 83 and 162 GPa.~\cite{degtyareva2005crystal} 
For Li$_2$S, we correctly
predict a transition from the antifluorite structure ($Fm\bar{3}m$)
 at ambient pressure to the anticotunnite structure at 13$\,$Gpa. At pressures
higher than 26$\,$GPa, a Ni$_2$In-type structure ($P6_3/mmc$), shown in Fig.~\ref{fig:elfs}, 
becomes stable; similar transition sequences are observed in other alkali-metal sulfides as Na$_2$S and K$_2$S, ~\cite{vegas2001reversible,vegas2002antifluorite} as well as in the closely related compound Li$_2$O. ~\cite{Lazicki_LiS_PRB2006} The $P6_3/mmc$ phase 
remains insulating up to 221$\,$GPa, where an insulator-to-metal transition takes place; we find that this
phase remains stable up to the highest pressure we calculated.

For pressures higher than 20 GPa, several new compositions are stabilized; the
relative structures are detailed in the supplementary material (SM), ~\cite{SM} 
together with band
 structure plots and results of DFPT calculations.
Figure~\ref{fig:elfs} only shows those  relevant to our discussion, decorated with isosurfaces of the
electronic localization function (ELF).
We start the discussion from the Li-rich side: The Li$_5$S composition becomes stable at 15 GPa, in an orthorombic $Cmmm$ structure,
with 12 atoms in the unit cell.
This is a very open and weakly metallic structure.
At $\sim$ 130 GPa, Li$_5$S transforms to a more densely packed $Immm$ phase,
shown in the first row of Fig.~\ref{fig:elfs}. The new phase shows signatures of interstitial charge localization
around the center of the tetragonal faces. 
The Li$_4$S stoichiometry is stabilized only at extreme pressures (P > 290 GPa); 
the lowest-enthalpy structure is trigonal, with
10 atoms in the unit cell. Due to the low symmetry
and poorly metallic behavior, we do not investigate this structure any further, but  
keep it in the convex hull because it has a strong infuence on the stability of other phases.

\lis\  is of particular interest, as it has the same stoichiometry as high-\tc\ \hs.
Its stability ranges from 20 GPa up to the highest pressure
investigated. 
 The lowest-enthalpy structure at 20 GPa, shown in the top left corner of Fig.~\ref{fig:elfs},
 has $Pm\bar{3}m$ space group.
Sulfur occupies the $1b$ positions at the center of the cube, 
and lithium  the $3d$ positions at the 
middle of the edges. Around the cube corners, 
there are large regions of empty space; the ELF isocontours show
that a substantial fraction of charge tends to localize in these regions.
Above 220 GPa, the simple cubic structure is destabilized 
towards an $I4/mmm$ variant, with three $f.u.$ in the unit cell,
in which three cubic cells are stacked along one of the cubic axes, with a small in-plane mismatch.
Except for the different stacking, the interatomic distances and interstitial 
charge localization are very similar to the $Pm\bar{3}m$ phase.
At 640 GPa, the simple cubic arrangement is finally destabilized towards
a completely different phase, with space group $Fm\bar{3}m$, 
shown in the bottom left corner of 
Fig.~\ref{fig:elfs}.
This structure, which has been reported at high pressures for Li$_3$N ~\cite{Lazicki_Li3N_PRL2005} is very closely packed.
In this case, 
the ELF shows that the valence charge, which can no longer
occupy the interstitial regions rearranges and 
Li forms strong bonds 
with its second nearest neighbor S along the $(100)$ direction, 
indicated by the red, dashed line in 
Fig.~\ref{fig:elfs}. 
As we will show in the following, the suppression of interstitial charge localization in $Fm\bar{3}m$-Li$_3$S is the reason this structure is the only high-\tc\ superconducting phase
of our study.
The two S-rich phases (LiS$_2$ and LiS$_3$) have very different 
structures and stability ranges. 
LiS$_3$ crystallizes in the same \bcc structure as \hs,
with S occupying the $6b$ Wyckoff positions of hydrogen, and Li the $2a$ 
of S;
this phase, shown in the middle of the first row in Fig.~\ref{fig:elfs},
is  thermodynamically stable only between 20 and 80$\,$GPa.
LiS$_2$ crystallizes in an $I4/mmm$ crystal structure, with two $f.u.$ -- middle of lower row in Fig.~\ref{fig:elfs} -- which is metallic and lies on the convex hull at pressures larger than 350 GPa.
Both phases are superconducting with moderate \tc's.

Our precedent study shows that there
are fundamentally three pressure regimes in the high-pressure phase
diagram of the Li-S system in Fig.~\ref{fig:PD}: (a) a low-pressure regime (P < 15$\,$GPa), where
Li$_2$S is the only stable composition; (b) an intermediate regime 
(P < 200$\,$GPa), where new stoichiometries are stabilized;
some of the new phases, such as LiS$_3$, disappear at higher pressures,  
while others remain; and (c) a high-pressure regime, where new 
phases appear again.
With the exception of Li$_2$S below 221$\,$GPa, we identified  all new phases 
as metallic,  which leaves us with an extremely large pool of
potential high-\tc\  conventional superconductors.

For a given crystal structure and chemical composition,
the superconducting \tc\  due to $ep$ interaction
can be estimated through the Mc-Millan-Allen-Dynes formula:
\begin{equation}\label{eq:tc}
  T_\text{c}=\frac{\omega_\text{log}}{1.2 k_B}\exp\left[-\frac{1.04(1+\lambda)}{\lambda-\mu^{*}(1+0.62\lambda)}\right]~.
\end{equation}
Here, $k_B$ is the Boltzmann constant and \mus\ 
is the Coulomb pseudopotential.
The $ep$ coupling constant 
$\lambda$ and the logarithmic average phonon frequency $\omega_\text{log}$ are obtained from the Eliashberg
spectral function for the $ep$ interaction $\alpha^2 F(\omega)$, 
calculated within DFPT:~\cite{QE_details}
\begin{equation}
 \alpha^2 F(\omega) = \frac{1}{N(E_F)} \sum \limits_{\mathbf{k} \mathbf{q},\nu} |g_{\mathbf{k},\mathbf{k}+\mathbf{q},\nu}|^2 \delta(\epsilon_\mathbf{k}) \delta(\epsilon_{\mathbf{k}+\mathbf{q}}) \delta(\omega-\omega_{\mathbf{q},\nu})~, \label{eq:alpha}
\end{equation}
as: $\lambda=2 \int d\omega \frac{\alpha^2 F(\omega)}{\omega}$; $\omega_\text{log}=\exp\left[\frac{2}{\lambda}\int \frac{d\omega}{\omega} \alpha^2 F(\omega) \ln(\omega) \right]$. In Eq.~\ref{eq:alpha}, 
 $N(E_F)$ is the density of states (DOS) at the Fermi level, $\omega_{\mathbf{q},\nu}$ is the phonon frequency
of mode $\nu$ and wavevector $\mathbf{q}$ and 
$|g_{\mathbf{k},\mathbf{k}+\mathbf{q},\nu}|$ is the electron-phonon matrix element between 
two electronic states with momenta $\mathbf{k}$ and $\mathbf{k+q}$ at the Fermi level.~\cite{Carbotte_RMP1990,AllenMitrovic1983}

Since $ep$ coupling calculations  in DFPT are computationally much more demanding than
 electronic structure calculations, we could not afford a full scan of the phase space at all pressures
and compositions. Instead, we selected two pressures, 100 and 500$\,$GPa,
representative of the intermediate and high-pressure regimes respectively,
and performed \tc{} calculations for all phases which are stable
at these pressures or in their immediate vicinity.

\begin{figure}[t]
\includegraphics[width=1.0\columnwidth]{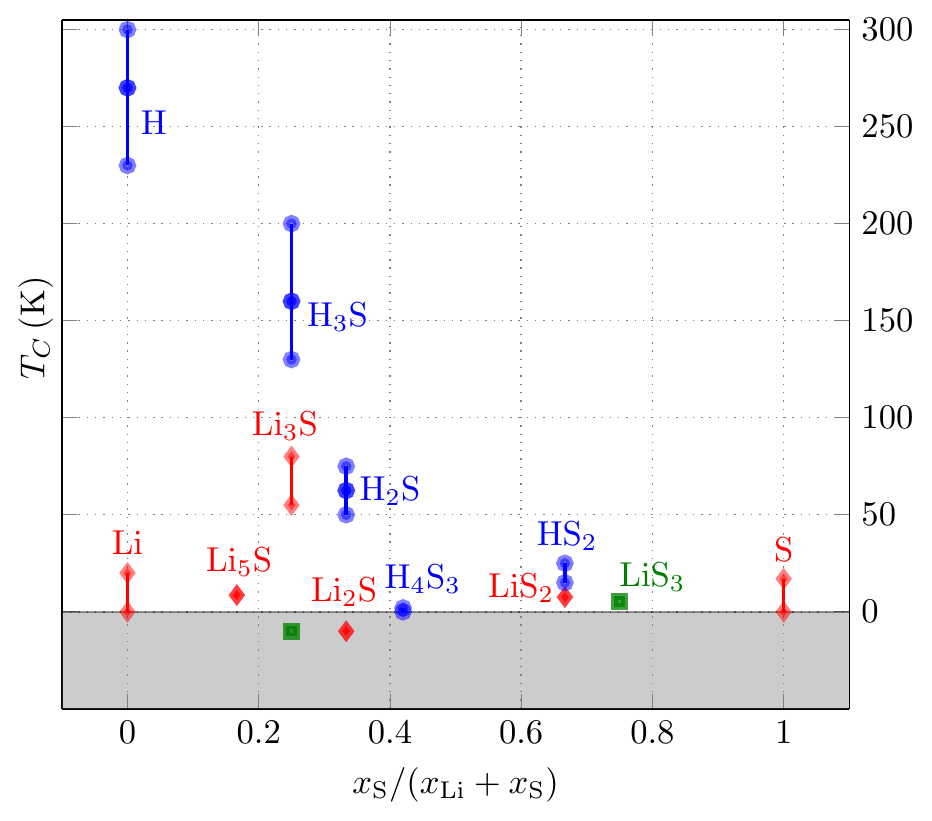}
\caption{\label{fig:tc}(Color online) 
Calculated \tc's of intermediate~(green) and high~(red) pressure Li-S phases. 
Blue symbols indicate values for the H-S system, taken from literature.~\cite{Li_dissociationHS_PRBR_2016,akashi_h3Smagneli_condmat2015}
The error bars show the variations of \tc\  due to pressure, when available.
Phases with no superconductivity are shown as negative values.
}
\end{figure}

\begin{table}
  \begin{center}
 \caption{\label{tab:lambda}
Superconducting properties of the metallic Li-S phases.
\tc's are estimated from Eq.~\ref{eq:tc}, with $\mu^*$=0.1.
Pressures are in GPa, $\omega_\text{log}$ and \tc's are in K;
 $\tilde{N}$(\ef) is the DOS at the Fermi level,
in st/Ry, divided by the number of atoms in the unit cell; $\eta$=$\lambda/\tilde{N}(E_F)$
is in Ry $\cdot$ atom.
Data for \hs\ are from Ref.~\onlinecite{FloresSanna_H3Se_EJPB2016}.}
\begin{ruledtabular}
    \begin{tabular}
{ l c  r  c  r  c  c  c  c}
comp. & $P$ & $\omega_\text{log}$ & $\lambda$ & \tc & $\tilde{N}$(\ef) & $\eta$
\\[1ex]
\hline
\lis\     ($Pm\bar{3}m$)  &100   &  754  & 0.08  & 0.0    & 0.62  &     0.13 \\[1ex]
LiS$_3$  ($Im\bar{3}m$)  &100   &  409   &  0.52 & 5.4    & 1.45  &     0.36 \\[1ex]
Li   ($P4_132$) & 500  &  546 &  0.40  & 2.2   & 0.25  & 1.64 \\[1ex]
Li$_5S$  ($Immm$)        & 500  &  420 &  0.53   & 8.6   & 0.48  & 1.10 \\[1ex]
\lis\     ($Pm\bar{3}m$)  &500   &  702   &  0.25  & 0.0   & 0.67  &     0.37 \\[1ex]
\lis\    ($Fm\bar{3}m$)   &500   &  773  &  1.43  & 80.0  & 1.67  &     0.85 \\[1ex]
\lis\    ($Fm\bar{3}m$) &600   &  826 &  1.01  & 55.9  & 1.30  & 0.78 \\[1ex]
Li$_2$S ($P6_3/mmc$)  & 500  &  374 &  0.22  & 0.0   & 0.27  & 0.85 \\[1ex]
LiS$_2$ ($I4/mmm$) & 500  &  494 &  0.54   & 7.6   & 1.35  & 0.40 \\[1ex]
\hline
\hs\ ($Im\bar{3}m$)       &200  & 1200 &   2.40  & 180  & 1.83 &     1.31 \\[1ex]
\lis$^{H}$ ($Fm\bar{3}m$) &500 & 1156 &   1.43   & 169 & 1.67 &     0.86 \\[1ex]
\end{tabular}
\end{ruledtabular}
\end{center}
\end{table}

The calculated values of \tc \ are plotted in Fig.~\ref{fig:tc}, as green (100$\,$GPa)
and red (500$\,$GPa) symbols.
Literature values for different sulfur hydrides from Ref.~\cite{Li_dissociationHS_PRBR_2016}
are shown as blue symbols on the same scale; the error bars indicate pressure variations
of \tc, when known.
To give a visual impression of
the presence or absence of superconductivity, \tc's smaller than 0.5$\,$K are 
shown as negative. Table  \ref{tab:lambda} reports the corresponding values
of $\omega_\text{log}$ and $\lambda$. Since we could not find any literature values, we calculated also data 
for the \textit{P}$4_132$ phase of Li at 500\,GPa.

Figure~\ref{fig:tc}  shows that only a few Li-S phases display 
a finite \tc, and a single phase -- i.e. the high-pressure $fcc$ phase of
Li$_3$S -- displays a critical temperature 
comparable to that of hydrides.
Furthermore, two of the phases with
 a finite \tc, LiS$_3$ and LiS$_2$, are S-rich phases,
in which the $ep$ coupling is dominated by the sulfur sublattice,
and thus not directly related to hydrides.
Other Li-rich phases, including elemental Li, exhibit \tc's lower than 20 K.
The contrast with the corresponding hydrides is striking: for some compositions
the differences in \tc \ are as large as two orders of magnitude. 
This suggests a fundamental difference between hydrogen- and lithium-rich compounds that we will investigate based on Eq.~\ref{eq:tc} and 
Tab.~\ref{tab:lambda}.

First, the higher atomic mass of lithium implies a smaller
prefactor $\omega_\text{log}$ in Eq.~\ref{eq:tc}:
to a first approximation, the reduction can be estimated as: 
%
%
$\sqrt{M_{Li}/M_{H}} \simeq 2.6$.
For most cases reported in Tab.~\ref{tab:lambda},  this is clearly not the dominating effect.
The single notable exception is $Fm\bar{3}m$-Li$_3$S, which is the only truly high-\tc\  phase 
identified in our study - its
\tc\  is 80\,K at 500\,GPa, and decreases to 55\,K in its stability range. If we take into 
account the mass effect, the \tc\ is comparable to that of \hs. 
To prove that, we performed a calculation in which we replaced the Li mass 
with that of hydrogen; this phase is indicated
as Li$_3$S$^{H}$ in the table. The calculated \tc\ is 170\,K, 
i.e., comparable to that of \hs. We want to note, however, that the pressure needed
to stabilize a high-\tc\  phase in this case
is almost three times larger as in the hydrides. We will discuss this point further in the following.

Table \ref{tab:lambda} shows that, with the exception of $Fm\bar{3}m$-Li$_3$S,
 where $\lambda \ge 1$,
all Li-S phases have small, or at best intermediate $ep$ coupling constants ($0.1 < \lambda < 0.55$).
%
The simplified Hopfield expression: 
$\lambda=\frac{N(E_F)I^2}{M\omega^2}$,
where $I$ is an average $ep$ 
matrix element and $M\omega^2$ is an average lattice force constant
permits to separate 
the $ep$ coupling into a purely electronic contribution given by the DOS
and a factor $\eta= \frac{I^2}{M\omega^2}$, related to the lattice.
Values of 
$\tilde{N}(E_F)$, i.e. DOS per atom, and $\eta$ for all Li-S phases in Fig.~\ref{fig:tc} 
 are reported in Tab.~\ref{tab:lambda}. 

First of all, we notice that in three high-pressure phases,
simple cubic Li, Li$_5$S and Li$_2$S, $\lambda$ is suppressed by an extremely low
$\tilde{N}(E_F)$. For Li and Li$_5$S, the poor 
metallic behavior
is a consequence of 2$s$-2$p$ hybridization; Li$_2$S is instead
a semiconducting phase which has metallized by band overlap, and its DOS 
is intrinsically low.
In the two sulfur-rich phases -- LiS$_2$
and LiS$_3$ -- the $ep$ coupling is moderate ($\lambda \sim 0.55$) and 
the DOS is sizable; due to the high sulfur content the characteristic phonon frequencies and \tc's are relatively low. 

Interestingly, for Li$_3$S 
we observe a striking difference between the simple cubic ($sc$)
$Pm\bar{3}m$ low-P and the 
$fcc$ $Fm\bar{3}m$  high-P structures:
The $sc$ phase has an extremely low $\lambda=0.08$ in its stability
range, which increases slightly at 500 GPa ($\lambda=0.25$), where it is still dynamically stable; both values
yield negligible \tc's. The high-\tc\  $fcc$ phase, instead, exhibits
a very high coupling ($\lambda=1.43$) at 500 GPa and $\lambda=1.01$ 
at higher pressures (600 GPa); the corresponding \tc's 
are large.

Table \ref{tab:lambda} shows that in this case, besides the DOS, there is a 
remarkable
difference in the lattice contribution to the $ep$ coupling, $\eta$, 
between the low and high-pressure phases.
This is due to the different nature of electronic
states involved in the superconducting pairing in the two structures.
In fact, the double-$\delta$ integral in  Eq.~\ref{eq:alpha} 
implies that the only electronic states, which give a finite contribution to the $ep$ coupling are those that are at $E_F$. If these states have
a large intrinsic coupling to phonons, as in covalently-bonded solids, 
$\eta$, and thus $\lambda$, are 
large.~\cite{SH_PRB-Mazin-2015,Heil-Boeri_PRB2015,boeri_diamond_PRL2004}
On the contrary, interstitial electrons,
which are localized in empty regions of the crystal structure,
couple very little to lattice vibrations, and hence 
$\eta$ and $\lambda$ will be low.

\begin{figure}[t]
\includegraphics[width=1.0\columnwidth,angle=0]{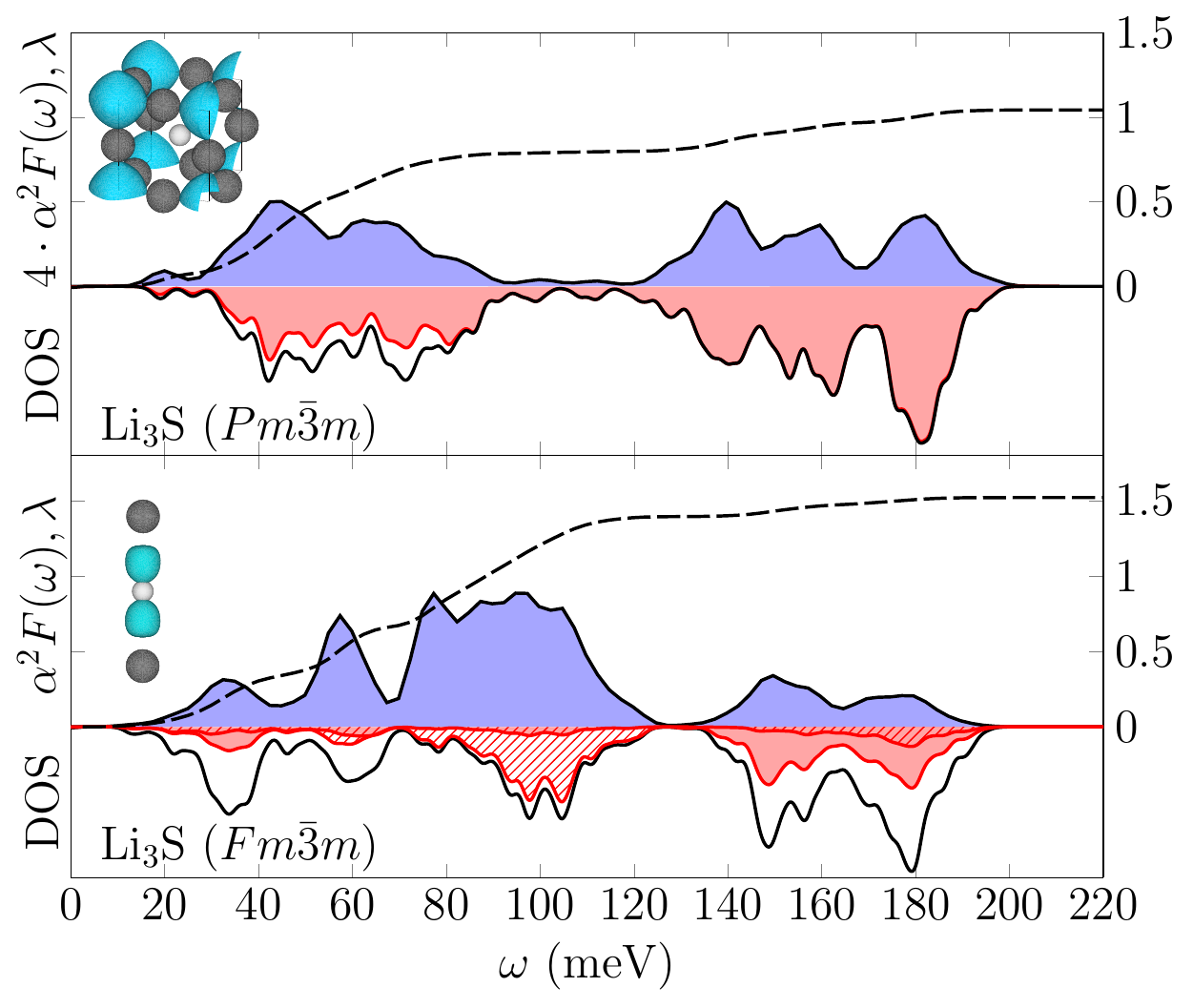}
\caption{\label{fig:alpha}
(Color online) Above: $\alpha^2 F(\omega)$ 
for $Pm\bar{3}m$--\lis\ (top) and $Fm\bar{3}m$--\lis\ (bottom)  at 500 GPa; 
the black, dashed lines show the frequency-integrated $ep$ 
coupling $\lambda(\omega)$. 
Note that the data for $Pm\bar{3}m$--\lis\ are multiplied by four.
Below: Phonon DOS's. Partial lithium contributions are shown in red; 
the dashed areas in $Fm\bar{3}m$--\lis\ indicate vibrations of the Li atoms which 
form bonds with S in the ($100$) direction -- red dashed lines in Fig.~\ref{fig:elfs}.
The insets show isocontours of the square of the wavefunction for the electronic bands that cross the Fermi level.}
\end{figure}

Figure~\ref{fig:alpha} illustrates how the superconducting 
properties of $Pm\bar{3}m$--
and $Fm\bar{3}m$--\lis\  differ due to matrix elements effects.
The two main panels 
show the Eliashberg spectral functions and partial phonon DOS's
calculated for both phases at 500\,GPa: 
the two spectra extend up to 180\,meV, but the intensity
and spectral distribution of the $ep$ coupling is crucially different.
Note that for better readability of the
figure, $\alpha^2F(\omega)$ and $\lambda$ of $Pm\bar{3}m$--\lis\ are multiplied by four.
 $Pm\bar{3}m$--\lis\ has an extremly uniform (and low) $ep$ coupling,
while $Fm\bar{3}m$--\lis\ shows a strong enhancement in the spectral region which corresponds to modes that distort the long Li-S
bonds in the $(100)$ direction.
In the small insets  we plot isocontours of 
the square of the wavefunctions for the 
electronic states at the Fermi level -- see SM for the definition.~\cite{SM}
In $Pm\bar{3}m$--\lis, these are localized in the interstitial region, i.e. 
around the corners of the cube. In $Fm\bar{3}m$--\lis, on the other hand,
they are localized along the
edges of the cube, i.e. along the long $(100)$ Li-S bonds, indicated by the red dashed lines in Fig.~\ref{fig:elfs}.
The different nature of the electronic states leads to a factor of $\sim 3$ increase in $\eta$; the difference in 
$\lambda$ is even larger. 

We thus find that interstitial charge localization due to avoided core overlap
can be a fundamental  limiting factor for conventional 
superconductivity.  This feature is very common in many alkali-metal-rich phases, including several new Li-S phases of this study,
indicated by dashed areas in in Fig.~\ref{fig:PD}.
When, as in $Pm\bar{3}m$--\lis,
 the electron count is such that interstitial 
charge localization involve electrons
at $E_F$, the $ep$ coupling is strongly suppressed. 

In conclusion, in this work we studied the thermodynamic stability and superconducting properties
of the Li-S system up to 700\,GPa, using methods for {\em ab-initio} crystal structure prediction
and linear response calculations of the $ep$ coupling. 
The calculated convex hulls show that several compositions besides the 
ambient pressure Li$_2$S are stabilized with increasing pressure. Most of these phases are
metallic, but exhibit no or low-\tc{} superconductivity.
We attribute this to two detrimental effects of core electrons in lithium: (i) an increased insulating behavior under pressure, due to hybridization between 2$s$ and 2$p$ electronic states; and (ii) interstitial charge localization due to avoided core overlap, which can bring states to the Fermi level that are intrinsically not coupled to lattice vibrations.
This is observed for example in \lis, where a \mbox{high-\tc}~(55-80\,K)
$fcc$ phase appears only at pressures high enough to 
stabilize closed packed structures (>\,600$\,$GPa).
Our study thus shows that  high-\tc \ superconductivity 
at megabar pressures can be attained in Li-rich compounds,
  similarly to hydrides,
but a general tendency to insulating behavior and  avoided core overlap will 
limit the possible range of pressures and dopings.

\begin{acknowledgments}
 The authors acknowledge computational resources from the {\em dCluster} of the Graz University of Technology and 
the VSC3 of the Vienna University of Technology.
\end{acknowledgments}

\bibliographystyle{apsrev4-1}
\bibliography{paper}

\begin{thebibliography}{70}%
\makeatletter
\providecommand \@ifxundefined [1]{%
 \@ifx{#1\undefined}
}%
\providecommand \@ifnum [1]{%
 \ifnum #1\expandafter \@firstoftwo
 \else \expandafter \@secondoftwo
 \fi
}%
\providecommand \@ifx [1]{%
 \ifx #1\expandafter \@firstoftwo
 \else \expandafter \@secondoftwo
 \fi
}%
\providecommand \natexlab [1]{#1}%
\providecommand \enquote  [1]{``#1''}%
\providecommand \bibnamefont  [1]{#1}%
\providecommand \bibfnamefont [1]{#1}%
\providecommand \citenamefont [1]{#1}%
\providecommand \href@noop [0]{\@secondoftwo}%
\providecommand \href [0]{\begingroup \@sanitize@url \@href}%
\providecommand \@href[1]{\@@startlink{#1}\@@href}%
\providecommand \@@href[1]{\endgroup#1\@@endlink}%
\providecommand \@sanitize@url [0]{\catcode `\\12\catcode `\$12\catcode
  `\&12\catcode `\#12\catcode `\^12\catcode `\_12\catcode `\%12\relax}%
\providecommand \@@startlink[1]{}%
\providecommand \@@endlink[0]{}%
\providecommand \url  [0]{\begingroup\@sanitize@url \@url }%
\providecommand \@url [1]{\endgroup\@href {#1}{\urlprefix }}%
\providecommand \urlprefix  [0]{URL }%
\providecommand \Eprint [0]{\href }%
\providecommand \doibase [0]{http://dx.doi.org/}%
\providecommand \selectlanguage [0]{\@gobble}%
\providecommand \bibinfo  [0]{\@secondoftwo}%
\providecommand \bibfield  [0]{\@secondoftwo}%
\providecommand \translation [1]{[#1]}%
\providecommand \BibitemOpen [0]{}%
\providecommand \bibitemStop [0]{}%
\providecommand \bibitemNoStop [0]{.\EOS\space}%
\providecommand \EOS [0]{\spacefactor3000\relax}%
\providecommand \BibitemShut  [1]{\csname bibitem#1\endcsname}%
\let\auto@bib@innerbib\@empty
\bibitem [{\citenamefont {Drozdov}\ \emph
  {et~al.}(2015{\natexlab{a}})\citenamefont {Drozdov}, \citenamefont {Eremets},
  \citenamefont {Troyan}, \citenamefont {Ksenofontov},\ and\ \citenamefont
  {Shylin}}]{DrozdovEremets_Nature2015}%
  \BibitemOpen
  \bibfield  {author} {\bibinfo {author} {\bibfnamefont {A.~P.}\ \bibnamefont
  {Drozdov}}, \bibinfo {author} {\bibfnamefont {M.~I.}\ \bibnamefont
  {Eremets}}, \bibinfo {author} {\bibfnamefont {I.~A.}\ \bibnamefont {Troyan}},
  \bibinfo {author} {\bibfnamefont {V.}~\bibnamefont {Ksenofontov}}, \ and\
  \bibinfo {author} {\bibfnamefont {S.~I.}\ \bibnamefont {Shylin}},\
  }\href@noop {} {\bibfield  {journal} {\bibinfo  {journal} {Nature}\ }\textbf
  {\bibinfo {volume} {000}},\ \bibinfo {pages} {2015/08/17/online} (\bibinfo
  {year} {2015}{\natexlab{a}})}\BibitemShut {NoStop}%
\bibitem [{\citenamefont {Troyan}\ \emph {et~al.}(2016)\citenamefont {Troyan},
  \citenamefont {Gavriliuk}, \citenamefont {R{\"u}ffer}, \citenamefont
  {Chumakov}, \citenamefont {Mironovich}, \citenamefont {Lyubutin},
  \citenamefont {Perekalin}, \citenamefont {Drozdov},\ and\ \citenamefont
  {Eremets}}]{Troyan_Science_2016}%
  \BibitemOpen
  \bibfield  {author} {\bibinfo {author} {\bibfnamefont {I.}~\bibnamefont
  {Troyan}}, \bibinfo {author} {\bibfnamefont {A.}~\bibnamefont {Gavriliuk}},
  \bibinfo {author} {\bibfnamefont {R.}~\bibnamefont {R{\"u}ffer}}, \bibinfo
  {author} {\bibfnamefont {A.}~\bibnamefont {Chumakov}}, \bibinfo {author}
  {\bibfnamefont {A.}~\bibnamefont {Mironovich}}, \bibinfo {author}
  {\bibfnamefont {I.}~\bibnamefont {Lyubutin}}, \bibinfo {author}
  {\bibfnamefont {D.}~\bibnamefont {Perekalin}}, \bibinfo {author}
  {\bibfnamefont {A.~P.}\ \bibnamefont {Drozdov}}, \ and\ \bibinfo {author}
  {\bibfnamefont {M.~I.}\ \bibnamefont {Eremets}},\ }\href {\doibase
  10.1126/science.aac8176} {\bibfield  {journal} {\bibinfo  {journal}
  {Science}\ }\textbf {\bibinfo {volume} {351}},\ \bibinfo {pages} {1303}
  (\bibinfo {year} {2016})},\ \Eprint
  {http://arxiv.org/abs/http://science.sciencemag.org/content/351/6279/1303.full.pdf}
  {http://science.sciencemag.org/content/351/6279/1303.full.pdf} \BibitemShut
  {NoStop}%
\bibitem [{\citenamefont {Duan}\ \emph {et~al.}(2014)\citenamefont {Duan},
  \citenamefont {Liu}, \citenamefont {Tian}, \citenamefont {Li}, \citenamefont
  {Huang}, \citenamefont {Zhao}, \citenamefont {Yu}, \citenamefont {Liu},
  \citenamefont {Tian},\ and\ \citenamefont {Cui}}]{Duan_SciRep2014}%
  \BibitemOpen
  \bibfield  {author} {\bibinfo {author} {\bibfnamefont {D.}~\bibnamefont
  {Duan}}, \bibinfo {author} {\bibfnamefont {Y.}~\bibnamefont {Liu}}, \bibinfo
  {author} {\bibfnamefont {F.}~\bibnamefont {Tian}}, \bibinfo {author}
  {\bibfnamefont {D.}~\bibnamefont {Li}}, \bibinfo {author} {\bibfnamefont
  {X.}~\bibnamefont {Huang}}, \bibinfo {author} {\bibfnamefont
  {Z.}~\bibnamefont {Zhao}}, \bibinfo {author} {\bibfnamefont {H.}~\bibnamefont
  {Yu}}, \bibinfo {author} {\bibfnamefont {B.}~\bibnamefont {Liu}}, \bibinfo
  {author} {\bibfnamefont {W.}~\bibnamefont {Tian}}, \ and\ \bibinfo {author}
  {\bibfnamefont {T.}~\bibnamefont {Cui}},\ }\href {\doibase
  http://dx.doi.org/10.1038/srep06968} {\bibfield  {journal} {\bibinfo
  {journal} {Sci. Rep.}\ }\textbf {\bibinfo {volume} {4}} (\bibinfo {year}
  {2014}),\ http://dx.doi.org/10.1038/srep06968}\BibitemShut {NoStop}%
\bibitem [{\citenamefont {Bernstein}\ \emph {et~al.}(2015)\citenamefont
  {Bernstein}, \citenamefont {Hellberg}, \citenamefont {Johannes},
  \citenamefont {Mazin},\ and\ \citenamefont {Mehl}}]{SH_PRB-Mazin-2015}%
  \BibitemOpen
  \bibfield  {author} {\bibinfo {author} {\bibfnamefont {N.}~\bibnamefont
  {Bernstein}}, \bibinfo {author} {\bibfnamefont {C.~S.}\ \bibnamefont
  {Hellberg}}, \bibinfo {author} {\bibfnamefont {M.~D.}\ \bibnamefont
  {Johannes}}, \bibinfo {author} {\bibfnamefont {I.~I.}\ \bibnamefont {Mazin}},
  \ and\ \bibinfo {author} {\bibfnamefont {M.~J.}\ \bibnamefont {Mehl}},\
  }\href {\doibase 10.1103/PhysRevB.91.060511} {\bibfield  {journal} {\bibinfo
  {journal} {Phys. Rev. B}\ }\textbf {\bibinfo {volume} {91}},\ \bibinfo
  {pages} {060511} (\bibinfo {year} {2015})}\BibitemShut {NoStop}%
\bibitem [{\citenamefont {Heil}\ and\ \citenamefont
  {Boeri}(2015)}]{Heil-Boeri_PRB2015}%
  \BibitemOpen
  \bibfield  {author} {\bibinfo {author} {\bibfnamefont {C.}~\bibnamefont
  {Heil}}\ and\ \bibinfo {author} {\bibfnamefont {L.}~\bibnamefont {Boeri}},\
  }\href {\doibase 10.1103/PhysRevB.92.060508} {\bibfield  {journal} {\bibinfo
  {journal} {Phys. Rev. B}\ }\textbf {\bibinfo {volume} {92}},\ \bibinfo
  {pages} {060508} (\bibinfo {year} {2015})}\BibitemShut {NoStop}%
\bibitem [{\citenamefont {{Flores-Livas, Jos\'e}}\ \emph
  {et~al.}(2016)\citenamefont {{Flores-Livas, Jos\'e}}, \citenamefont {{Sanna,
  Antonio}},\ and\ \citenamefont {{Gross, E.
  K.U.}}}]{FloresSanna_H3Se_EJPB2016}%
  \BibitemOpen
  \bibfield  {author} {\bibinfo {author} {\bibnamefont {{Flores-Livas,
  Jos\'e}}}, \bibinfo {author} {\bibnamefont {{Sanna, Antonio}}}, \ and\
  \bibinfo {author} {\bibnamefont {{Gross, E. K.U.}}},\ }\href {\doibase
  10.1140/epjb/e2016-70020-0} {\bibfield  {journal} {\bibinfo  {journal} {Eur.
  Phys. J. B}\ }\textbf {\bibinfo {volume} {89}},\ \bibinfo {pages} {63}
  (\bibinfo {year} {2016})}\BibitemShut {NoStop}%
\bibitem [{\citenamefont {Quan}\ and\ \citenamefont
  {Pickett}(2016)}]{quan_impact_2015}%
  \BibitemOpen
  \bibfield  {author} {\bibinfo {author} {\bibfnamefont {Y.}~\bibnamefont
  {Quan}}\ and\ \bibinfo {author} {\bibfnamefont {W.~E.}\ \bibnamefont
  {Pickett}},\ }\href {\doibase 10.1103/PhysRevB.93.104526} {\bibfield
  {journal} {\bibinfo  {journal} {Phys. Rev. B}\ }\textbf {\bibinfo {volume}
  {93}},\ \bibinfo {pages} {104526} (\bibinfo {year} {2016})}\BibitemShut
  {NoStop}%
\bibitem [{\citenamefont {Li}\ \emph {et~al.}(2014)\citenamefont {Li},
  \citenamefont {Hao}, \citenamefont {Liu}, \citenamefont {Li},\ and\
  \citenamefont {Ma}}]{LiYanmingMa_JCP2014}%
  \BibitemOpen
  \bibfield  {author} {\bibinfo {author} {\bibfnamefont {Y.}~\bibnamefont
  {Li}}, \bibinfo {author} {\bibfnamefont {J.}~\bibnamefont {Hao}}, \bibinfo
  {author} {\bibfnamefont {H.}~\bibnamefont {Liu}}, \bibinfo {author}
  {\bibfnamefont {Y.}~\bibnamefont {Li}}, \ and\ \bibinfo {author}
  {\bibfnamefont {Y.}~\bibnamefont {Ma}},\ }\href {\doibase
  http://dx.doi.org/10.1063/1.4874158} {\bibfield  {journal} {\bibinfo
  {journal} {The Journal of Chemical Physics}\ }\textbf {\bibinfo {volume}
  {140}},\  (\bibinfo {year} {2014})}\BibitemShut {NoStop}%
\bibitem [{\citenamefont {Papaconstantopoulos}\ \emph
  {et~al.}(2015)\citenamefont {Papaconstantopoulos}, \citenamefont {Klein},
  \citenamefont {Mehl},\ and\ \citenamefont
  {Pickett}}]{Papaconstantopoulos_H3S_PRB2015}%
  \BibitemOpen
  \bibfield  {author} {\bibinfo {author} {\bibfnamefont {D.~A.}\ \bibnamefont
  {Papaconstantopoulos}}, \bibinfo {author} {\bibfnamefont {B.~M.}\
  \bibnamefont {Klein}}, \bibinfo {author} {\bibfnamefont {M.~J.}\ \bibnamefont
  {Mehl}}, \ and\ \bibinfo {author} {\bibfnamefont {W.~E.}\ \bibnamefont
  {Pickett}},\ }\href {\doibase 10.1103/PhysRevB.91.184511} {\bibfield
  {journal} {\bibinfo  {journal} {Phys. Rev. B}\ }\textbf {\bibinfo {volume}
  {91}},\ \bibinfo {pages} {184511} (\bibinfo {year} {2015})}\BibitemShut
  {NoStop}%
\bibitem [{\citenamefont {Luciano}\ \emph {et~al.}(2015)\citenamefont
  {Luciano}, \citenamefont {Cappelluti},\ and\ \citenamefont
  {Pietronero}}]{Ortenzi_TB_2015}%
  \BibitemOpen
  \bibfield  {author} {\bibinfo {author} {\bibfnamefont {O.}~\bibnamefont
  {Luciano}}, \bibinfo {author} {\bibfnamefont {E.}~\bibnamefont {Cappelluti}},
  \ and\ \bibinfo {author} {\bibfnamefont {L.}~\bibnamefont {Pietronero}},\
  }\href {http://arxiv.org/abs/1511.04304} {\  (\bibinfo {year} {2015})},\
  \bibinfo {note} {arXiv: 1511.04304}\BibitemShut {NoStop}%
\bibitem [{\citenamefont {Errea}\ \emph {et~al.}(2015)\citenamefont {Errea},
  \citenamefont {Calandra}, \citenamefont {Pickard}, \citenamefont {Nelson},
  \citenamefont {Needs}, \citenamefont {Li}, \citenamefont {Liu}, \citenamefont
  {Zhang}, \citenamefont {Ma},\ and\ \citenamefont
  {Mauri}}]{Errea_anhaPRL2015}%
  \BibitemOpen
  \bibfield  {author} {\bibinfo {author} {\bibfnamefont {I.}~\bibnamefont
  {Errea}}, \bibinfo {author} {\bibfnamefont {M.}~\bibnamefont {Calandra}},
  \bibinfo {author} {\bibfnamefont {C.~J.}\ \bibnamefont {Pickard}}, \bibinfo
  {author} {\bibfnamefont {J.}~\bibnamefont {Nelson}}, \bibinfo {author}
  {\bibfnamefont {R.~J.}\ \bibnamefont {Needs}}, \bibinfo {author}
  {\bibfnamefont {Y.}~\bibnamefont {Li}}, \bibinfo {author} {\bibfnamefont
  {H.}~\bibnamefont {Liu}}, \bibinfo {author} {\bibfnamefont {Y.}~\bibnamefont
  {Zhang}}, \bibinfo {author} {\bibfnamefont {Y.}~\bibnamefont {Ma}}, \ and\
  \bibinfo {author} {\bibfnamefont {F.}~\bibnamefont {Mauri}},\ }\href
  {\doibase 10.1103/PhysRevLett.114.157004} {\bibfield  {journal} {\bibinfo
  {journal} {Phys. Rev. Lett.}\ }\textbf {\bibinfo {volume} {114}},\ \bibinfo
  {pages} {157004} (\bibinfo {year} {2015})}\BibitemShut {NoStop}%
\bibitem [{\citenamefont {Drozdov}\ \emph
  {et~al.}(2015{\natexlab{b}})\citenamefont {Drozdov}, \citenamefont
  {Eremets},\ and\ \citenamefont {Troyan}}]{Drozdov_ph3_arxiv2015}%
  \BibitemOpen
  \bibfield  {author} {\bibinfo {author} {\bibfnamefont {A.}~\bibnamefont
  {Drozdov}}, \bibinfo {author} {\bibfnamefont {M.~I.}\ \bibnamefont
  {Eremets}}, \ and\ \bibinfo {author} {\bibfnamefont {I.~A.}\ \bibnamefont
  {Troyan}},\ }\href@noop {} {\bibfield  {journal} {\bibinfo  {journal} {ArXiv
  e-prints}\ } (\bibinfo {year} {2015}{\natexlab{b}})},\ \Eprint
  {http://arxiv.org/abs/1508.06224} {arXiv:1508.06224 [cond-mat.supr-con]}
  \BibitemShut {NoStop}%
\bibitem [{\citenamefont {Shamp}\ \emph {et~al.}(2016)\citenamefont {Shamp},
  \citenamefont {Terpstra}, \citenamefont {Bi}, \citenamefont {Falls},
  \citenamefont {Avery},\ and\ \citenamefont
  {Zurek}}]{shamp_decomposition_2015}%
  \BibitemOpen
  \bibfield  {author} {\bibinfo {author} {\bibfnamefont {A.}~\bibnamefont
  {Shamp}}, \bibinfo {author} {\bibfnamefont {T.}~\bibnamefont {Terpstra}},
  \bibinfo {author} {\bibfnamefont {T.}~\bibnamefont {Bi}}, \bibinfo {author}
  {\bibfnamefont {Z.}~\bibnamefont {Falls}}, \bibinfo {author} {\bibfnamefont
  {P.}~\bibnamefont {Avery}}, \ and\ \bibinfo {author} {\bibfnamefont
  {E.}~\bibnamefont {Zurek}},\ }\href {\doibase 10.1021/jacs.5b10180}
  {\bibfield  {journal} {\bibinfo  {journal} {Journal of the American Chemical
  Society}\ }\textbf {\bibinfo {volume} {138}},\ \bibinfo {pages} {1884}
  (\bibinfo {year} {2016})},\ \bibinfo {note} {pMID: 26777416},\ \Eprint
  {http://arxiv.org/abs/http://dx.doi.org/10.1021/jacs.5b10180}
  {http://dx.doi.org/10.1021/jacs.5b10180} \BibitemShut {NoStop}%
\bibitem [{\citenamefont {Flores-Livas}\ \emph {et~al.}(2016)\citenamefont
  {Flores-Livas}, \citenamefont {Amsler}, \citenamefont {Heil}, \citenamefont
  {Sanna}, \citenamefont {Boeri}, \citenamefont {Profeta}, \citenamefont
  {Wolverton}, \citenamefont {Goedecker},\ and\ \citenamefont
  {Gross}}]{Flores_PH3_PRBR2016}%
  \BibitemOpen
  \bibfield  {author} {\bibinfo {author} {\bibfnamefont {J.~A.}\ \bibnamefont
  {Flores-Livas}}, \bibinfo {author} {\bibfnamefont {M.}~\bibnamefont
  {Amsler}}, \bibinfo {author} {\bibfnamefont {C.}~\bibnamefont {Heil}},
  \bibinfo {author} {\bibfnamefont {A.}~\bibnamefont {Sanna}}, \bibinfo
  {author} {\bibfnamefont {L.}~\bibnamefont {Boeri}}, \bibinfo {author}
  {\bibfnamefont {G.}~\bibnamefont {Profeta}}, \bibinfo {author} {\bibfnamefont
  {C.}~\bibnamefont {Wolverton}}, \bibinfo {author} {\bibfnamefont
  {S.}~\bibnamefont {Goedecker}}, \ and\ \bibinfo {author} {\bibfnamefont
  {E.~K.~U.}\ \bibnamefont {Gross}},\ }\href {\doibase
  10.1103/PhysRevB.93.020508} {\bibfield  {journal} {\bibinfo  {journal} {Phys.
  Rev. B}\ }\textbf {\bibinfo {volume} {93}},\ \bibinfo {pages} {020508}
  (\bibinfo {year} {2016})}\BibitemShut {NoStop}%
\bibitem [{\citenamefont {Fu}\ \emph {et~al.}(2016)\citenamefont {Fu},
  \citenamefont {Du}, \citenamefont {Zhang}, \citenamefont {Peng},
  \citenamefont {Zhang}, \citenamefont {Pickard}, \citenamefont {Needs},
  \citenamefont {Singh}, \citenamefont {Zheng},\ and\ \citenamefont
  {Ma}}]{Fu_Ma_pnictogenH_2016}%
  \BibitemOpen
  \bibfield  {author} {\bibinfo {author} {\bibfnamefont {Y.}~\bibnamefont
  {Fu}}, \bibinfo {author} {\bibfnamefont {X.}~\bibnamefont {Du}}, \bibinfo
  {author} {\bibfnamefont {L.}~\bibnamefont {Zhang}}, \bibinfo {author}
  {\bibfnamefont {F.}~\bibnamefont {Peng}}, \bibinfo {author} {\bibfnamefont
  {M.}~\bibnamefont {Zhang}}, \bibinfo {author} {\bibfnamefont {C.~J.}\
  \bibnamefont {Pickard}}, \bibinfo {author} {\bibfnamefont {R.~J.}\
  \bibnamefont {Needs}}, \bibinfo {author} {\bibfnamefont {D.~J.}\ \bibnamefont
  {Singh}}, \bibinfo {author} {\bibfnamefont {W.}~\bibnamefont {Zheng}}, \ and\
  \bibinfo {author} {\bibfnamefont {Y.}~\bibnamefont {Ma}},\ }\href {\doibase
  10.1021/acs.chemmater.5b04638} {\bibfield  {journal} {\bibinfo  {journal}
  {Chemistry of Materials}\ }\textbf {\bibinfo {volume} {28}},\ \bibinfo
  {pages} {1746} (\bibinfo {year} {2016})},\ \Eprint
  {http://arxiv.org/abs/http://dx.doi.org/10.1021/acs.chemmater.5b04638}
  {http://dx.doi.org/10.1021/acs.chemmater.5b04638} \BibitemShut {NoStop}%
\bibitem [{\citenamefont {Ashcroft}(1968)}]{Ashcroft_PRL1968}%
  \BibitemOpen
  \bibfield  {author} {\bibinfo {author} {\bibfnamefont {N.}~\bibnamefont
  {Ashcroft}},\ }\href {\doibase 10.1103/PhysRevLett.21.1748} {\bibfield
  {journal} {\bibinfo  {journal} {Phys. Rev. Lett.}\ }\textbf {\bibinfo
  {volume} {21}},\ \bibinfo {pages} {1748} (\bibinfo {year}
  {1968})}\BibitemShut {NoStop}%
\bibitem [{\citenamefont {Cudazzo}\ \emph {et~al.}(2008)\citenamefont
  {Cudazzo}, \citenamefont {Profeta}, \citenamefont {Sanna}, \citenamefont
  {Floris}, \citenamefont {Continenza}, \citenamefont {Massidda},\ and\
  \citenamefont {Gross}}]{Cudazzo_PRL2008}%
  \BibitemOpen
  \bibfield  {author} {\bibinfo {author} {\bibfnamefont {P.}~\bibnamefont
  {Cudazzo}}, \bibinfo {author} {\bibfnamefont {G.}~\bibnamefont {Profeta}},
  \bibinfo {author} {\bibfnamefont {A.}~\bibnamefont {Sanna}}, \bibinfo
  {author} {\bibfnamefont {A.}~\bibnamefont {Floris}}, \bibinfo {author}
  {\bibfnamefont {A.}~\bibnamefont {Continenza}}, \bibinfo {author}
  {\bibfnamefont {S.}~\bibnamefont {Massidda}}, \ and\ \bibinfo {author}
  {\bibfnamefont {E.}~\bibnamefont {Gross}},\ }\href {\doibase
  10.1103/PhysRevLett.100.257001} {\bibfield  {journal} {\bibinfo  {journal}
  {Phys. Rev. Lett.}\ }\textbf {\bibinfo {volume} {100}},\ \bibinfo {pages}
  {257001} (\bibinfo {year} {2008})}\BibitemShut {NoStop}%
\bibitem [{\citenamefont {McMahon}\ and\ \citenamefont
  {Ceperley}(2011)}]{mcmahon_high_2011}%
  \BibitemOpen
  \bibfield  {author} {\bibinfo {author} {\bibfnamefont {J.~M.}\ \bibnamefont
  {McMahon}}\ and\ \bibinfo {author} {\bibfnamefont {D.~M.}\ \bibnamefont
  {Ceperley}},\ }\href {\doibase 10.1103/PhysRevB.84.144515} {\bibfield
  {journal} {\bibinfo  {journal} {Phys. Rev. B}\ }\textbf {\bibinfo {volume}
  {84}},\ \bibinfo {pages} {144515} (\bibinfo {year} {2011})}\BibitemShut
  {NoStop}%
\bibitem [{\citenamefont {Borinaga}\ \emph {et~al.}(2016)\citenamefont
  {Borinaga}, \citenamefont {Errea}, \citenamefont {Calandra}, \citenamefont
  {Mauri},\ and\ \citenamefont {Bergara}}]{Borinaga_H_arxiv2016}%
  \BibitemOpen
  \bibfield  {author} {\bibinfo {author} {\bibfnamefont {M.}~\bibnamefont
  {Borinaga}}, \bibinfo {author} {\bibfnamefont {I.}~\bibnamefont {Errea}},
  \bibinfo {author} {\bibfnamefont {M.}~\bibnamefont {Calandra}}, \bibinfo
  {author} {\bibfnamefont {F.}~\bibnamefont {Mauri}}, \ and\ \bibinfo {author}
  {\bibfnamefont {A.}~\bibnamefont {Bergara}},\ }\href@noop {} {\bibfield
  {journal} {\bibinfo  {journal} {ArXiv e-prints}\ } (\bibinfo {year}
  {2016})},\ \Eprint {http://arxiv.org/abs/1602.06877} {arXiv:1602.06877
  [cond-mat.supr-con]} \BibitemShut {NoStop}%
\bibitem [{\citenamefont {Szcz\c{e}\`{s}niak}\ and\ \citenamefont
  {Jarosik}(2009)}]{Szcz_superconducting_2009}%
  \BibitemOpen
  \bibfield  {author} {\bibinfo {author} {\bibfnamefont {R.}~\bibnamefont
  {Szcz\c{e}\`{s}niak}}\ and\ \bibinfo {author} {\bibfnamefont
  {M.}~\bibnamefont {Jarosik}},\ }\href {\doibase
  http://dx.doi.org/10.1016/j.ssc.2009.08.019} {\bibfield  {journal} {\bibinfo
  {journal} {Solid State Communications}\ }\textbf {\bibinfo {volume} {149}},\
  \bibinfo {pages} {2053} (\bibinfo {year} {2009})}\BibitemShut {NoStop}%
\bibitem [{\citenamefont {M.I.~Eremets}(2016)}]{Eremets_H_arxiv2016}%
  \BibitemOpen
  \bibfield  {author} {\bibinfo {author} {\bibfnamefont {A.~D.}\ \bibnamefont
  {M.I.~Eremets}, \bibfnamefont {I.A.~Troyan}},\ }\href@noop {} {\bibfield
  {journal} {\bibinfo  {journal} {ArXiv e-prints}\ } (\bibinfo {year}
  {2016})},\ \Eprint {http://arxiv.org/abs/1601.04479} {arXiv:1601.04479
  [cond-mat.supr-con]} \BibitemShut {NoStop}%
\bibitem [{\citenamefont {Dalladay-Simpson}(2016)}]{Dalladay_H_Nature2016}%
  \BibitemOpen
  \bibfield  {author} {\bibinfo {author} {\bibfnamefont {E.~G.}\ \bibnamefont
  {Dalladay-Simpson}, \bibfnamefont {R.~T.~Howie}},\ }\href@noop {} {\bibfield
  {journal} {\bibinfo  {journal} {Nature}\ }\textbf {\bibinfo {volume} {529}},\
  \bibinfo {pages} {63} (\bibinfo {year} {2016})}\BibitemShut {NoStop}%
\bibitem [{\citenamefont {Ashcroft}(2004)}]{Ashcroft_PRL2004}%
  \BibitemOpen
  \bibfield  {author} {\bibinfo {author} {\bibfnamefont {N.}~\bibnamefont
  {Ashcroft}},\ }\href {\doibase 10.1103/PhysRevLett.92.187002} {\bibfield
  {journal} {\bibinfo  {journal} {Phys. Rev. Lett.}\ }\textbf {\bibinfo
  {volume} {92}},\ \bibinfo {pages} {187002} (\bibinfo {year}
  {2004})}\BibitemShut {NoStop}%
\bibitem [{\citenamefont {Tse}\ \emph {et~al.}(2007)\citenamefont {Tse},
  \citenamefont {Yao},\ and\ \citenamefont {Tanaka}}]{tse_novel_2007}%
  \BibitemOpen
  \bibfield  {author} {\bibinfo {author} {\bibfnamefont {J.~S.}\ \bibnamefont
  {Tse}}, \bibinfo {author} {\bibfnamefont {Y.}~\bibnamefont {Yao}}, \ and\
  \bibinfo {author} {\bibfnamefont {K.}~\bibnamefont {Tanaka}},\ }\href
  {\doibase 10.1103/PhysRevLett.98.117004} {\bibfield  {journal} {\bibinfo
  {journal} {Phy. Rev. Lett.}\ }\textbf {\bibinfo {volume} {98}},\ \bibinfo
  {pages} {117004} (\bibinfo {year} {2007})}\BibitemShut {NoStop}%
\bibitem [{\citenamefont {Gao}\ \emph {et~al.}(2010)\citenamefont {Gao},
  \citenamefont {Oganov}, \citenamefont {Li}, \citenamefont {Li}, \citenamefont
  {Wang}, \citenamefont {Cui}, \citenamefont {Ma}, \citenamefont {Bergara},
  \citenamefont {Lyakhov}, \citenamefont {Iitaka},\ and\ \citenamefont
  {Zou}}]{gao_high-pressure_2010}%
  \BibitemOpen
  \bibfield  {author} {\bibinfo {author} {\bibfnamefont {G.}~\bibnamefont
  {Gao}}, \bibinfo {author} {\bibfnamefont {A.~R.}\ \bibnamefont {Oganov}},
  \bibinfo {author} {\bibfnamefont {P.}~\bibnamefont {Li}}, \bibinfo {author}
  {\bibfnamefont {Z.}~\bibnamefont {Li}}, \bibinfo {author} {\bibfnamefont
  {H.}~\bibnamefont {Wang}}, \bibinfo {author} {\bibfnamefont {T.}~\bibnamefont
  {Cui}}, \bibinfo {author} {\bibfnamefont {Y.}~\bibnamefont {Ma}}, \bibinfo
  {author} {\bibfnamefont {A.}~\bibnamefont {Bergara}}, \bibinfo {author}
  {\bibfnamefont {A.~O.}\ \bibnamefont {Lyakhov}}, \bibinfo {author}
  {\bibfnamefont {T.}~\bibnamefont {Iitaka}}, \ and\ \bibinfo {author}
  {\bibfnamefont {G.}~\bibnamefont {Zou}},\ }\href {\doibase
  10.1073/pnas.0908342107} {\bibfield  {journal} {\bibinfo  {journal}
  {Proceedings of the National Academy of Sciences}\ }\textbf {\bibinfo
  {volume} {107}},\ \bibinfo {pages} {1317} (\bibinfo {year}
  {2010})}\BibitemShut {NoStop}%
\bibitem [{\citenamefont {Kim}\ \emph {et~al.}(2010)\citenamefont {Kim},
  \citenamefont {Scheicher}, \citenamefont {Mao}, \citenamefont {Kang},\ and\
  \citenamefont {Ahuja}}]{kim_general_2010}%
  \BibitemOpen
  \bibfield  {author} {\bibinfo {author} {\bibfnamefont {D.~Y.}\ \bibnamefont
  {Kim}}, \bibinfo {author} {\bibfnamefont {R.~H.}\ \bibnamefont {Scheicher}},
  \bibinfo {author} {\bibfnamefont {H.-k.}\ \bibnamefont {Mao}}, \bibinfo
  {author} {\bibfnamefont {T.~W.}\ \bibnamefont {Kang}}, \ and\ \bibinfo
  {author} {\bibfnamefont {R.}~\bibnamefont {Ahuja}},\ }\href {\doibase
  10.1073/pnas.0914462107} {\bibfield  {journal} {\bibinfo  {journal} {PNAS}\
  }\textbf {\bibinfo {volume} {107}},\ \bibinfo {pages} {2793} (\bibinfo {year}
  {2010})}\BibitemShut {NoStop}%
\bibitem [{\citenamefont {Yao}\ and\ \citenamefont
  {Klug}(2010)}]{Yao_PNAS2010}%
  \BibitemOpen
  \bibfield  {author} {\bibinfo {author} {\bibfnamefont {Y.}~\bibnamefont
  {Yao}}\ and\ \bibinfo {author} {\bibfnamefont {D.~D.}\ \bibnamefont {Klug}},\
  }\href {\doibase 10.1073/pnas.1006508107} {\bibfield  {journal} {\bibinfo
  {journal} {Proceedings of the National Academy of Sciences}\ }\textbf
  {\bibinfo {volume} {107}},\ \bibinfo {pages} {20893} (\bibinfo {year}
  {2010})}\BibitemShut {NoStop}%
\bibitem [{\citenamefont {Flores-Livas}\ \emph {et~al.}(2012)\citenamefont
  {Flores-Livas}, \citenamefont {Amsler}, \citenamefont {Lenosky},
  \citenamefont {Lehtovaara}, \citenamefont {Botti}, \citenamefont {Marques},\
  and\ \citenamefont {Goedecker}}]{Disilane_JAFL}%
  \BibitemOpen
  \bibfield  {author} {\bibinfo {author} {\bibfnamefont {J.~A.}\ \bibnamefont
  {Flores-Livas}}, \bibinfo {author} {\bibfnamefont {M.}~\bibnamefont
  {Amsler}}, \bibinfo {author} {\bibfnamefont {T.~J.}\ \bibnamefont {Lenosky}},
  \bibinfo {author} {\bibfnamefont {L.}~\bibnamefont {Lehtovaara}}, \bibinfo
  {author} {\bibfnamefont {S.}~\bibnamefont {Botti}}, \bibinfo {author}
  {\bibfnamefont {M.~A.~L.}\ \bibnamefont {Marques}}, \ and\ \bibinfo {author}
  {\bibfnamefont {S.}~\bibnamefont {Goedecker}},\ }\href {\doibase
  10.1103/PhysRevLett.108.117004} {\bibfield  {journal} {\bibinfo  {journal}
  {Phys. Rev. Lett.}\ }\textbf {\bibinfo {volume} {108}},\ \bibinfo {pages}
  {117004} (\bibinfo {year} {2012})}\BibitemShut {NoStop}%
\bibitem [{\citenamefont {Oganov}\ \emph {et~al.}()\citenamefont {Oganov},
  \citenamefont {Glass}, \citenamefont {Lyakhov}, \citenamefont {Stokes},
  \citenamefont {Zhu}, \citenamefont {Agarwal}, \citenamefont {Dong},
  \citenamefont {Pertierra}, \citenamefont {Raza}, \citenamefont {Salvado}
  \emph {et~al.}}]{oganovuniversal}%
  \BibitemOpen
  \bibfield  {author} {\bibinfo {author} {\bibfnamefont {A.}~\bibnamefont
  {Oganov}}, \bibinfo {author} {\bibfnamefont {C.}~\bibnamefont {Glass}},
  \bibinfo {author} {\bibfnamefont {A.}~\bibnamefont {Lyakhov}}, \bibinfo
  {author} {\bibfnamefont {H.}~\bibnamefont {Stokes}}, \bibinfo {author}
  {\bibfnamefont {Q.}~\bibnamefont {Zhu}}, \bibinfo {author} {\bibfnamefont
  {R.}~\bibnamefont {Agarwal}}, \bibinfo {author} {\bibfnamefont
  {X.}~\bibnamefont {Dong}}, \bibinfo {author} {\bibfnamefont {P.}~\bibnamefont
  {Pertierra}}, \bibinfo {author} {\bibfnamefont {Z.}~\bibnamefont {Raza}},
  \bibinfo {author} {\bibfnamefont {M.}~\bibnamefont {Salvado}},  \emph
  {et~al.},\ }\href {"http://han.ess.sunysb.edu/uspex_manual/uspex_manual.pdf"}
  {\ }\BibitemShut {NoStop}%
\bibitem [{QE_()}]{QE_details}%
  \BibitemOpen
  \href@noop {} {}\bibinfo {note} {For selected phases we calculated the $ep$
  properties using DFPT, as implemented in \textsc{Quantum
  espresso}.~\cite{QE-2009}. We employed PBE ultrasoft pseudopotentials with
  semicore states in valence for lithium and sulfur from the standard
  \textsc{Quantum espresso} distribution, with energy cutoffs of 80 and 800\;Ry
  for wavefunctions and charge density, respectively. We used $8^3$
  $\mathbf{k}$- and $\mathbf{q}$-points meshes for reciprocal space integration
  for electron and phonon states in the self-consistent calculations, and up to
  38$^3$ $\mathbf{k}$-points for the $ep$ matrix elements.}\BibitemShut {Stop}%
\bibitem [{\citenamefont {Giannozzi}\ \emph {et~al.}(2009)\citenamefont
  {Giannozzi}, \citenamefont {Baroni}, \citenamefont {Bonini}, \citenamefont
  {Calandra}, \citenamefont {Car}, \citenamefont {Cavazzoni}, \citenamefont
  {Ceresoli}, \citenamefont {Chiarotti}, \citenamefont {Cococcioni},
  \citenamefont {Dabo}, \citenamefont {{Dal Corso}}, \citenamefont
  {de~Gironcoli}, \citenamefont {Fabris}, \citenamefont {Fratesi},
  \citenamefont {Gebauer}, \citenamefont {Gerstmann}, \citenamefont
  {Gougoussis}, \citenamefont {Kokalj}, \citenamefont {Lazzeri}, \citenamefont
  {Martin-Samos}, \citenamefont {Marzari}, \citenamefont {Mauri}, \citenamefont
  {Mazzarello}, \citenamefont {Paolini}, \citenamefont {Pasquarello},
  \citenamefont {Paulatto}, \citenamefont {Sbraccia}, \citenamefont {Scandolo},
  \citenamefont {Sclauzero}, \citenamefont {Seitsonen}, \citenamefont
  {Smogunov}, \citenamefont {Umari},\ and\ \citenamefont
  {Wentzcovitch}}]{QE-2009}%
  \BibitemOpen
  \bibfield  {author} {\bibinfo {author} {\bibfnamefont {P.}~\bibnamefont
  {Giannozzi}}, \bibinfo {author} {\bibfnamefont {S.}~\bibnamefont {Baroni}},
  \bibinfo {author} {\bibfnamefont {N.}~\bibnamefont {Bonini}}, \bibinfo
  {author} {\bibfnamefont {M.}~\bibnamefont {Calandra}}, \bibinfo {author}
  {\bibfnamefont {R.}~\bibnamefont {Car}}, \bibinfo {author} {\bibfnamefont
  {C.}~\bibnamefont {Cavazzoni}}, \bibinfo {author} {\bibfnamefont
  {D.}~\bibnamefont {Ceresoli}}, \bibinfo {author} {\bibfnamefont {G.~L.}\
  \bibnamefont {Chiarotti}}, \bibinfo {author} {\bibfnamefont {M.}~\bibnamefont
  {Cococcioni}}, \bibinfo {author} {\bibfnamefont {I.}~\bibnamefont {Dabo}},
  \bibinfo {author} {\bibfnamefont {A.}~\bibnamefont {{Dal Corso}}}, \bibinfo
  {author} {\bibfnamefont {S.}~\bibnamefont {de~Gironcoli}}, \bibinfo {author}
  {\bibfnamefont {S.}~\bibnamefont {Fabris}}, \bibinfo {author} {\bibfnamefont
  {G.}~\bibnamefont {Fratesi}}, \bibinfo {author} {\bibfnamefont
  {R.}~\bibnamefont {Gebauer}}, \bibinfo {author} {\bibfnamefont
  {U.}~\bibnamefont {Gerstmann}}, \bibinfo {author} {\bibfnamefont
  {C.}~\bibnamefont {Gougoussis}}, \bibinfo {author} {\bibfnamefont
  {A.}~\bibnamefont {Kokalj}}, \bibinfo {author} {\bibfnamefont
  {M.}~\bibnamefont {Lazzeri}}, \bibinfo {author} {\bibfnamefont
  {L.}~\bibnamefont {Martin-Samos}}, \bibinfo {author} {\bibfnamefont
  {N.}~\bibnamefont {Marzari}}, \bibinfo {author} {\bibfnamefont
  {F.}~\bibnamefont {Mauri}}, \bibinfo {author} {\bibfnamefont
  {R.}~\bibnamefont {Mazzarello}}, \bibinfo {author} {\bibfnamefont
  {S.}~\bibnamefont {Paolini}}, \bibinfo {author} {\bibfnamefont
  {A.}~\bibnamefont {Pasquarello}}, \bibinfo {author} {\bibfnamefont
  {L.}~\bibnamefont {Paulatto}}, \bibinfo {author} {\bibfnamefont
  {C.}~\bibnamefont {Sbraccia}}, \bibinfo {author} {\bibfnamefont
  {S.}~\bibnamefont {Scandolo}}, \bibinfo {author} {\bibfnamefont
  {G.}~\bibnamefont {Sclauzero}}, \bibinfo {author} {\bibfnamefont {A.~P.}\
  \bibnamefont {Seitsonen}}, \bibinfo {author} {\bibfnamefont {A.}~\bibnamefont
  {Smogunov}}, \bibinfo {author} {\bibfnamefont {P.}~\bibnamefont {Umari}}, \
  and\ \bibinfo {author} {\bibfnamefont {R.~M.}\ \bibnamefont {Wentzcovitch}},\
  }\href {http://www.quantum-espresso.org} {\bibfield  {journal} {\bibinfo
  {journal} {Journal of Physics: Condensed Matter}\ }\textbf {\bibinfo {volume}
  {21}},\ \bibinfo {pages} {395502 (19pp)} (\bibinfo {year}
  {2009})}\BibitemShut {NoStop}%
\bibitem [{\citenamefont {Grzechnik}\ \emph {et~al.}(2000)\citenamefont
  {Grzechnik}, \citenamefont {Vegas}, \citenamefont {Syassen}, \citenamefont
  {Loa}, \citenamefont {Hanfland},\ and\ \citenamefont
  {Jansen}}]{grzechnik2000reversible}%
  \BibitemOpen
  \bibfield  {author} {\bibinfo {author} {\bibfnamefont {A.}~\bibnamefont
  {Grzechnik}}, \bibinfo {author} {\bibfnamefont {A.}~\bibnamefont {Vegas}},
  \bibinfo {author} {\bibfnamefont {K.}~\bibnamefont {Syassen}}, \bibinfo
  {author} {\bibfnamefont {I.}~\bibnamefont {Loa}}, \bibinfo {author}
  {\bibfnamefont {M.}~\bibnamefont {Hanfland}}, \ and\ \bibinfo {author}
  {\bibfnamefont {M.}~\bibnamefont {Jansen}},\ }\href
  {http://www.sciencedirect.com/science/article/pii/S0022459600989023}
  {\bibfield  {journal} {\bibinfo  {journal} {Journal of Solid State
  Chemistry}\ }\textbf {\bibinfo {volume} {154}},\ \bibinfo {pages} {603}
  (\bibinfo {year} {2000})}\BibitemShut {NoStop}%
\bibitem [{\citenamefont {Lazicki}\ \emph {et~al.}(2006)\citenamefont
  {Lazicki}, \citenamefont {Yoo}, \citenamefont {Evans},\ and\ \citenamefont
  {Pickett}}]{Lazicki_LiS_PRB2006}%
  \BibitemOpen
  \bibfield  {author} {\bibinfo {author} {\bibfnamefont {A.}~\bibnamefont
  {Lazicki}}, \bibinfo {author} {\bibfnamefont {C.-S.}\ \bibnamefont {Yoo}},
  \bibinfo {author} {\bibfnamefont {W.~J.}\ \bibnamefont {Evans}}, \ and\
  \bibinfo {author} {\bibfnamefont {W.~E.}\ \bibnamefont {Pickett}},\ }\href
  {\doibase 10.1103/PhysRevB.73.184120} {\bibfield  {journal} {\bibinfo
  {journal} {Phys. Rev. B}\ }\textbf {\bibinfo {volume} {73}},\ \bibinfo
  {pages} {184120} (\bibinfo {year} {2006})}\BibitemShut {NoStop}%
\bibitem [{\citenamefont {Naumov}\ \emph {et~al.}(2015)\citenamefont {Naumov},
  \citenamefont {Hemley}, \citenamefont {Hoffmann},\ and\ \citenamefont
  {Ashcroft}}]{naumov2015chemical}%
  \BibitemOpen
  \bibfield  {author} {\bibinfo {author} {\bibfnamefont {I.~I.}\ \bibnamefont
  {Naumov}}, \bibinfo {author} {\bibfnamefont {R.~J.}\ \bibnamefont {Hemley}},
  \bibinfo {author} {\bibfnamefont {R.}~\bibnamefont {Hoffmann}}, \ and\
  \bibinfo {author} {\bibfnamefont {N.}~\bibnamefont {Ashcroft}},\ }\href
  {http://scitation.aip.org/content/aip/journal/jcp/143/6/10.1063/1.4928076}
  {\bibfield  {journal} {\bibinfo  {journal} {The Journal of chemical physics}\
  }\textbf {\bibinfo {volume} {143}},\ \bibinfo {pages} {064702} (\bibinfo
  {year} {2015})}\BibitemShut {NoStop}%
\bibitem [{\citenamefont {Hanfland}\ \emph {et~al.}(2000)\citenamefont
  {Hanfland}, \citenamefont {Syassen}, \citenamefont {Christensen},\ and\
  \citenamefont {Novikov}}]{hanfland2000new}%
  \BibitemOpen
  \bibfield  {author} {\bibinfo {author} {\bibfnamefont {M.}~\bibnamefont
  {Hanfland}}, \bibinfo {author} {\bibfnamefont {K.}~\bibnamefont {Syassen}},
  \bibinfo {author} {\bibfnamefont {N.}~\bibnamefont {Christensen}}, \ and\
  \bibinfo {author} {\bibfnamefont {D.}~\bibnamefont {Novikov}},\ }\href
  {http://www.nature.com/nature/journal/v408/n6809/abs/408174a0.html}
  {\bibfield  {journal} {\bibinfo  {journal} {Nature}\ }\textbf {\bibinfo
  {volume} {408}},\ \bibinfo {pages} {174} (\bibinfo {year}
  {2000})}\BibitemShut {NoStop}%
\bibitem [{\citenamefont {Shi}\ and\ \citenamefont
  {Papaconstantopoulos}(2006)}]{shi2006theoretical}%
  \BibitemOpen
  \bibfield  {author} {\bibinfo {author} {\bibfnamefont {L.}~\bibnamefont
  {Shi}}\ and\ \bibinfo {author} {\bibfnamefont {D.~A.}\ \bibnamefont
  {Papaconstantopoulos}},\ }\href
  {http://journals.aps.org/prb/abstract/10.1103/PhysRevB.73.184516} {\bibfield
  {journal} {\bibinfo  {journal} {Physical Review B}\ }\textbf {\bibinfo
  {volume} {73}},\ \bibinfo {pages} {184516} (\bibinfo {year}
  {2006})}\BibitemShut {NoStop}%
\bibitem [{\citenamefont {Christensen}\ and\ \citenamefont
  {Novikov}(2006)}]{christensen2006calculated}%
  \BibitemOpen
  \bibfield  {author} {\bibinfo {author} {\bibfnamefont {N.}~\bibnamefont
  {Christensen}}\ and\ \bibinfo {author} {\bibfnamefont {D.}~\bibnamefont
  {Novikov}},\ }\href
  {http://journals.aps.org/prb/abstract/10.1103/PhysRevB.73.224508} {\bibfield
  {journal} {\bibinfo  {journal} {Physical Review B}\ }\textbf {\bibinfo
  {volume} {73}},\ \bibinfo {pages} {224508} (\bibinfo {year}
  {2006})}\BibitemShut {NoStop}%
\bibitem [{\citenamefont {Marqu\'es}\ \emph
  {et~al.}(2011{\natexlab{a}})\citenamefont {Marqu\'es}, \citenamefont
  {McMahon}, \citenamefont {Gregoryanz}, \citenamefont {Hanfland},
  \citenamefont {Guillaume}, \citenamefont {Pickard}, \citenamefont {Ackland},\
  and\ \citenamefont {Nelmes}}]{marques_Lidense_PRL2011}%
  \BibitemOpen
  \bibfield  {author} {\bibinfo {author} {\bibfnamefont {M.}~\bibnamefont
  {Marqu\'es}}, \bibinfo {author} {\bibfnamefont {M.~I.}\ \bibnamefont
  {McMahon}}, \bibinfo {author} {\bibfnamefont {E.}~\bibnamefont {Gregoryanz}},
  \bibinfo {author} {\bibfnamefont {M.}~\bibnamefont {Hanfland}}, \bibinfo
  {author} {\bibfnamefont {C.~L.}\ \bibnamefont {Guillaume}}, \bibinfo {author}
  {\bibfnamefont {C.~J.}\ \bibnamefont {Pickard}}, \bibinfo {author}
  {\bibfnamefont {G.~J.}\ \bibnamefont {Ackland}}, \ and\ \bibinfo {author}
  {\bibfnamefont {R.~J.}\ \bibnamefont {Nelmes}},\ }\href {\doibase
  10.1103/PhysRevLett.106.095502} {\bibfield  {journal} {\bibinfo  {journal}
  {Phys. Rev. Lett.}\ }\textbf {\bibinfo {volume} {106}},\ \bibinfo {pages}
  {095502} (\bibinfo {year} {2011}{\natexlab{a}})}\BibitemShut {NoStop}%
\bibitem [{\citenamefont {Rousseau}\ and\ \citenamefont
  {Ashcroft}(2008)}]{Ashcroft_interstitial_PRL2008}%
  \BibitemOpen
  \bibfield  {author} {\bibinfo {author} {\bibfnamefont {B.}~\bibnamefont
  {Rousseau}}\ and\ \bibinfo {author} {\bibfnamefont {N.~W.}\ \bibnamefont
  {Ashcroft}},\ }\href {\doibase 10.1103/PhysRevLett.101.046407} {\bibfield
  {journal} {\bibinfo  {journal} {Phys. Rev. Lett.}\ }\textbf {\bibinfo
  {volume} {101}},\ \bibinfo {pages} {046407} (\bibinfo {year}
  {2008})}\BibitemShut {NoStop}%
\bibitem [{\citenamefont {Marqu\'es}\ \emph {et~al.}(2009)\citenamefont
  {Marqu\'es}, \citenamefont {Ackland}, \citenamefont {Lundegaard},
  \citenamefont {Stinton}, \citenamefont {Nelmes}, \citenamefont {McMahon},\
  and\ \citenamefont {Contreras-Garc\'{\i}a}}]{PhysRevLett.103.115501}%
  \BibitemOpen
  \bibfield  {author} {\bibinfo {author} {\bibfnamefont {M.}~\bibnamefont
  {Marqu\'es}}, \bibinfo {author} {\bibfnamefont {G.~J.}\ \bibnamefont
  {Ackland}}, \bibinfo {author} {\bibfnamefont {L.~F.}\ \bibnamefont
  {Lundegaard}}, \bibinfo {author} {\bibfnamefont {G.}~\bibnamefont {Stinton}},
  \bibinfo {author} {\bibfnamefont {R.~J.}\ \bibnamefont {Nelmes}}, \bibinfo
  {author} {\bibfnamefont {M.~I.}\ \bibnamefont {McMahon}}, \ and\ \bibinfo
  {author} {\bibfnamefont {J.}~\bibnamefont {Contreras-Garc\'{\i}a}},\ }\href
  {\doibase 10.1103/PhysRevLett.103.115501} {\bibfield  {journal} {\bibinfo
  {journal} {Phys. Rev. Lett.}\ }\textbf {\bibinfo {volume} {103}},\ \bibinfo
  {pages} {115501} (\bibinfo {year} {2009})}\BibitemShut {NoStop}%
\bibitem [{\citenamefont {Gatti}\ \emph {et~al.}(2010)\citenamefont {Gatti},
  \citenamefont {Tokatly},\ and\ \citenamefont
  {Rubio}}]{PhysRevLett.104.216404}%
  \BibitemOpen
  \bibfield  {author} {\bibinfo {author} {\bibfnamefont {M.}~\bibnamefont
  {Gatti}}, \bibinfo {author} {\bibfnamefont {I.~V.}\ \bibnamefont {Tokatly}},
  \ and\ \bibinfo {author} {\bibfnamefont {A.}~\bibnamefont {Rubio}},\ }\href
  {\doibase 10.1103/PhysRevLett.104.216404} {\bibfield  {journal} {\bibinfo
  {journal} {Phys. Rev. Lett.}\ }\textbf {\bibinfo {volume} {104}},\ \bibinfo
  {pages} {216404} (\bibinfo {year} {2010})}\BibitemShut {NoStop}%
\bibitem [{\citenamefont {Marqu\'es}\ \emph
  {et~al.}(2011{\natexlab{b}})\citenamefont {Marqu\'es}, \citenamefont
  {McMahon}, \citenamefont {Gregoryanz}, \citenamefont {Hanfland},
  \citenamefont {Guillaume}, \citenamefont {Pickard}, \citenamefont {Ackland},\
  and\ \citenamefont {Nelmes}}]{PhysRevLett.106.095502}%
  \BibitemOpen
  \bibfield  {author} {\bibinfo {author} {\bibfnamefont {M.}~\bibnamefont
  {Marqu\'es}}, \bibinfo {author} {\bibfnamefont {M.~I.}\ \bibnamefont
  {McMahon}}, \bibinfo {author} {\bibfnamefont {E.}~\bibnamefont {Gregoryanz}},
  \bibinfo {author} {\bibfnamefont {M.}~\bibnamefont {Hanfland}}, \bibinfo
  {author} {\bibfnamefont {C.~L.}\ \bibnamefont {Guillaume}}, \bibinfo {author}
  {\bibfnamefont {C.~J.}\ \bibnamefont {Pickard}}, \bibinfo {author}
  {\bibfnamefont {G.~J.}\ \bibnamefont {Ackland}}, \ and\ \bibinfo {author}
  {\bibfnamefont {R.~J.}\ \bibnamefont {Nelmes}},\ }\href {\doibase
  10.1103/PhysRevLett.106.095502} {\bibfield  {journal} {\bibinfo  {journal}
  {Phys. Rev. Lett.}\ }\textbf {\bibinfo {volume} {106}},\ \bibinfo {pages}
  {095502} (\bibinfo {year} {2011}{\natexlab{b}})}\BibitemShut {NoStop}%
\bibitem [{\citenamefont {Xie}\ \emph {et~al.}(2010)\citenamefont {Xie},
  \citenamefont {Oganov},\ and\ \citenamefont {Ma}}]{Oganov_CaLi2_PRL2010}%
  \BibitemOpen
  \bibfield  {author} {\bibinfo {author} {\bibfnamefont {Y.}~\bibnamefont
  {Xie}}, \bibinfo {author} {\bibfnamefont {A.~R.}\ \bibnamefont {Oganov}}, \
  and\ \bibinfo {author} {\bibfnamefont {Y.}~\bibnamefont {Ma}},\ }\href
  {\doibase 10.1103/PhysRevLett.104.177005} {\bibfield  {journal} {\bibinfo
  {journal} {Phys. Rev. Lett.}\ }\textbf {\bibinfo {volume} {104}},\ \bibinfo
  {pages} {177005} (\bibinfo {year} {2010})}\BibitemShut {NoStop}%
\bibitem [{\citenamefont {Pickard}\ and\ \citenamefont
  {Needs}(2007)}]{pickard_structure_2007}%
  \BibitemOpen
  \bibfield  {author} {\bibinfo {author} {\bibfnamefont {C.~J.}\ \bibnamefont
  {Pickard}}\ and\ \bibinfo {author} {\bibfnamefont {R.~J.}\ \bibnamefont
  {Needs}},\ }\href {\doibase 10.1038/nphys625} {\bibfield  {journal} {\bibinfo
   {journal} {Nat Phys}\ }\textbf {\bibinfo {volume} {3}},\ \bibinfo {pages}
  {473} (\bibinfo {year} {2007})}\BibitemShut {NoStop}%
\bibitem [{\citenamefont {Yao}\ \emph {et~al.}(2009)\citenamefont {Yao},
  \citenamefont {Tse}, \citenamefont {Tanaka}, \citenamefont {Marsiglio},\ and\
  \citenamefont {Ma}}]{yao2009superconductivity}%
  \BibitemOpen
  \bibfield  {author} {\bibinfo {author} {\bibfnamefont {Y.}~\bibnamefont
  {Yao}}, \bibinfo {author} {\bibfnamefont {J.}~\bibnamefont {Tse}}, \bibinfo
  {author} {\bibfnamefont {K.}~\bibnamefont {Tanaka}}, \bibinfo {author}
  {\bibfnamefont {F.}~\bibnamefont {Marsiglio}}, \ and\ \bibinfo {author}
  {\bibfnamefont {Y.}~\bibnamefont {Ma}},\ }\href
  {http://journals.aps.org/prb/abstract/10.1103/PhysRevB.79.054524} {\bibfield
  {journal} {\bibinfo  {journal} {Physical Review B}\ }\textbf {\bibinfo
  {volume} {79}},\ \bibinfo {pages} {054524} (\bibinfo {year}
  {2009})}\BibitemShut {NoStop}%
\bibitem [{\citenamefont {Profeta}\ \emph {et~al.}(2006)\citenamefont
  {Profeta}, \citenamefont {Franchini}, \citenamefont {Lathiotakis},
  \citenamefont {Floris}, \citenamefont {Sanna}, \citenamefont {Marques},
  \citenamefont {L\"uders}, \citenamefont {Massidda}, \citenamefont {Gross},\
  and\ \citenamefont {Continenza}}]{Profeta_LiKAl_PRL2006}%
  \BibitemOpen
  \bibfield  {author} {\bibinfo {author} {\bibfnamefont {G.}~\bibnamefont
  {Profeta}}, \bibinfo {author} {\bibfnamefont {C.}~\bibnamefont {Franchini}},
  \bibinfo {author} {\bibfnamefont {N.}~\bibnamefont {Lathiotakis}}, \bibinfo
  {author} {\bibfnamefont {A.}~\bibnamefont {Floris}}, \bibinfo {author}
  {\bibfnamefont {A.}~\bibnamefont {Sanna}}, \bibinfo {author} {\bibfnamefont
  {M.~A.~L.}\ \bibnamefont {Marques}}, \bibinfo {author} {\bibfnamefont
  {M.}~\bibnamefont {L\"uders}}, \bibinfo {author} {\bibfnamefont
  {S.}~\bibnamefont {Massidda}}, \bibinfo {author} {\bibfnamefont {E.~K.~U.}\
  \bibnamefont {Gross}}, \ and\ \bibinfo {author} {\bibfnamefont
  {A.}~\bibnamefont {Continenza}},\ }\href {\doibase
  10.1103/PhysRevLett.96.047003} {\bibfield  {journal} {\bibinfo  {journal}
  {Phys. Rev. Lett.}\ }\textbf {\bibinfo {volume} {96}},\ \bibinfo {pages}
  {047003} (\bibinfo {year} {2006})}\BibitemShut {NoStop}%
\bibitem [{\citenamefont {Murnaghan}(1944)}]{murnaghan1944compressibility}%
  \BibitemOpen
  \bibfield  {author} {\bibinfo {author} {\bibfnamefont {F.}~\bibnamefont
  {Murnaghan}},\ }\href {http://www.pnas.org/content/30/9/244.short} {\bibfield
   {journal} {\bibinfo  {journal} {Proceedings of the National Academy of
  Sciences}\ }\textbf {\bibinfo {volume} {30}},\ \bibinfo {pages} {244}
  (\bibinfo {year} {1944})}\BibitemShut {NoStop}%
\bibitem [{USP()}]{USPEX_details}%
  \BibitemOpen
  \href@noop {} {}\bibinfo {note} {We explored the Li-S system using the
  technique of \textit{ab-initio} evolutionary crystal structure prediction as
  implemented in the
  \textsc{USPEX}-package.~\cite{oganov2006crystal,lyakhov2013new,oganov2011evolutionary}
  The underlying structural relaxations were performed using the \textsc{VASP}
  code,~\cite{kresse1993ab,kresse1996efficiency} within the generalized
  gradient approximation.~\cite{perdew1996generalized} We used the all-electron
  projector-augmented wave
  method.~\cite{kresse1999ultrasoft,blochl1994projector} To avoid core overlap
  at high pressures in Li, we treated the 1s and 2s electrons as valence.
  Individual structures generated by the evolutionary algorithms were relaxed
  with increasing precision in a 5-step procedure; the energies were finally
  recalculated with increasing convergence criteria to ensure a correct ranking
  of the structure.}\BibitemShut {Stop}%
\bibitem [{PD_()}]{PD_details}%
  \BibitemOpen
  \href@noop {} {}\bibinfo {note} {For all evolutionary runs, we limited
  trapping in local minima by repeating simulations at selected pressures and
  using the antiseeds-technique described in Ref.~\cite{oganovuniversal}. To
  ensure a reliable ranking of the structures, we performed extensive
  convergence checks of the total energy and forces with respect to cut-off
  energy and k-point resolution; for variable and fixed composition runs we
  used energy cut-offs up to 700 up to 800$\,$eV, and a k-point resolution of
  0.06\,$2 \pi /{\text \AA}$ to obtain a first rough approximation of the
  enthalpies; for the final enthalpy vs. pressure curves, we increased these
  values up to 1100 to 1200$\,$eV and Monkhorst-Pack k-point meshes for
  sampling the Brilloin-zone with resolution of 0.03\,$2 \pi /{\text \AA}$,
  which ensured a convergence of forces up to 1\,meV/atom.}\BibitemShut {Stop}%
\bibitem [{\citenamefont {Ma}\ \emph {et~al.}(2008)\citenamefont {Ma},
  \citenamefont {Oganov},\ and\ \citenamefont {Xie}}]{ma2008high}%
  \BibitemOpen
  \bibfield  {author} {\bibinfo {author} {\bibfnamefont {Y.}~\bibnamefont
  {Ma}}, \bibinfo {author} {\bibfnamefont {A.~R.}\ \bibnamefont {Oganov}}, \
  and\ \bibinfo {author} {\bibfnamefont {Y.}~\bibnamefont {Xie}},\ }\href
  {http://journals.aps.org/prb/abstract/10.1103/PhysRevB.78.014102} {\bibfield
  {journal} {\bibinfo  {journal} {Physical Review B}\ }\textbf {\bibinfo
  {volume} {78}},\ \bibinfo {pages} {014102} (\bibinfo {year}
  {2008})}\BibitemShut {NoStop}%
\bibitem [{\citenamefont {Oganov}\ and\ \citenamefont
  {Glass}(2006{\natexlab{a}})}]{oganov_sulphur}%
  \BibitemOpen
  \bibfield  {author} {\bibinfo {author} {\bibfnamefont {A.~R.}\ \bibnamefont
  {Oganov}}\ and\ \bibinfo {author} {\bibfnamefont {C.~W.}\ \bibnamefont
  {Glass}},\ }\href {\doibase http://dx.doi.org/10.1063/1.2210932} {\bibfield
  {journal} {\bibinfo  {journal} {The Journal of Chemical Physics}\ }\textbf
  {\bibinfo {volume} {124}},\ \bibinfo {eid} {244704} (\bibinfo {year}
  {2006}{\natexlab{a}}),\ http://dx.doi.org/10.1063/1.2210932}\BibitemShut
  {NoStop}%
\bibitem [{\citenamefont {Zakharov}\ and\ \citenamefont
  {Cohen}(1995)}]{zakharov1995theory}%
  \BibitemOpen
  \bibfield  {author} {\bibinfo {author} {\bibfnamefont {O.}~\bibnamefont
  {Zakharov}}\ and\ \bibinfo {author} {\bibfnamefont {M.~L.}\ \bibnamefont
  {Cohen}},\ }\href
  {http://journals.aps.org/prb/abstract/10.1103/PhysRevB.52.12572} {\bibfield
  {journal} {\bibinfo  {journal} {Physical Review B}\ }\textbf {\bibinfo
  {volume} {52}},\ \bibinfo {pages} {12572} (\bibinfo {year}
  {1995})}\BibitemShut {NoStop}%
\bibitem [{\citenamefont {Degtyareva}\ \emph {et~al.}(2005)\citenamefont
  {Degtyareva}, \citenamefont {Gregoryanz}, \citenamefont {Somayazulu},
  \citenamefont {Mao},\ and\ \citenamefont {Hemley}}]{degtyareva2005crystal}%
  \BibitemOpen
  \bibfield  {author} {\bibinfo {author} {\bibfnamefont {O.}~\bibnamefont
  {Degtyareva}}, \bibinfo {author} {\bibfnamefont {E.}~\bibnamefont
  {Gregoryanz}}, \bibinfo {author} {\bibfnamefont {M.}~\bibnamefont
  {Somayazulu}}, \bibinfo {author} {\bibfnamefont {H.-k.}\ \bibnamefont {Mao}},
  \ and\ \bibinfo {author} {\bibfnamefont {R.~J.}\ \bibnamefont {Hemley}},\
  }\href {http://journals.aps.org/prb/abstract/10.1103/PhysRevB.71.214104}
  {\bibfield  {journal} {\bibinfo  {journal} {Physical Review B}\ }\textbf
  {\bibinfo {volume} {71}},\ \bibinfo {pages} {214104} (\bibinfo {year}
  {2005})}\BibitemShut {NoStop}%
\bibitem [{\citenamefont {Vegas}\ \emph {et~al.}(2001)\citenamefont {Vegas},
  \citenamefont {Grzechnik}, \citenamefont {Syassen}, \citenamefont {Loa},
  \citenamefont {Hanfland},\ and\ \citenamefont
  {Jansen}}]{vegas2001reversible}%
  \BibitemOpen
  \bibfield  {author} {\bibinfo {author} {\bibfnamefont {A.}~\bibnamefont
  {Vegas}}, \bibinfo {author} {\bibfnamefont {A.}~\bibnamefont {Grzechnik}},
  \bibinfo {author} {\bibfnamefont {K.}~\bibnamefont {Syassen}}, \bibinfo
  {author} {\bibfnamefont {I.}~\bibnamefont {Loa}}, \bibinfo {author}
  {\bibfnamefont {M.}~\bibnamefont {Hanfland}}, \ and\ \bibinfo {author}
  {\bibfnamefont {M.}~\bibnamefont {Jansen}},\ }\href
  {http://scripts.iucr.org/cgi-bin/paper?S0108768100016621} {\bibfield
  {journal} {\bibinfo  {journal} {Acta Crystallographica Section B: Structural
  Science}\ }\textbf {\bibinfo {volume} {57}},\ \bibinfo {pages} {151}
  (\bibinfo {year} {2001})}\BibitemShut {NoStop}%
\bibitem [{\citenamefont {Vegas}\ \emph {et~al.}(2002)\citenamefont {Vegas},
  \citenamefont {Grzechnik}, \citenamefont {Hanfland}, \citenamefont
  {M{\"u}hle},\ and\ \citenamefont {Jansen}}]{vegas2002antifluorite}%
  \BibitemOpen
  \bibfield  {author} {\bibinfo {author} {\bibfnamefont {A.}~\bibnamefont
  {Vegas}}, \bibinfo {author} {\bibfnamefont {A.}~\bibnamefont {Grzechnik}},
  \bibinfo {author} {\bibfnamefont {M.}~\bibnamefont {Hanfland}}, \bibinfo
  {author} {\bibfnamefont {C.}~\bibnamefont {M{\"u}hle}}, \ and\ \bibinfo
  {author} {\bibfnamefont {M.}~\bibnamefont {Jansen}},\ }\href
  {http://www.sciencedirect.com/science/article/pii/S1293255802013602}
  {\bibfield  {journal} {\bibinfo  {journal} {Solid state sciences}\ }\textbf
  {\bibinfo {volume} {4}},\ \bibinfo {pages} {1077} (\bibinfo {year}
  {2002})}\BibitemShut {NoStop}%
\bibitem [{SM()}]{SM}%
  \BibitemOpen
  \href@noop {} {}\bibinfo {note} {Supplementary Material for this article is
  available online under...}\BibitemShut {Stop}%
\bibitem [{\citenamefont {Lazicki}\ \emph {et~al.}(2005)\citenamefont
  {Lazicki}, \citenamefont {Maddox}, \citenamefont {Evans}, \citenamefont
  {Yoo}, \citenamefont {McMahan}, \citenamefont {Pickett}, \citenamefont
  {Scalettar}, \citenamefont {Hu},\ and\ \citenamefont
  {Chow}}]{Lazicki_Li3N_PRL2005}%
  \BibitemOpen
  \bibfield  {author} {\bibinfo {author} {\bibfnamefont {A.}~\bibnamefont
  {Lazicki}}, \bibinfo {author} {\bibfnamefont {B.}~\bibnamefont {Maddox}},
  \bibinfo {author} {\bibfnamefont {W.~J.}\ \bibnamefont {Evans}}, \bibinfo
  {author} {\bibfnamefont {C.-S.}\ \bibnamefont {Yoo}}, \bibinfo {author}
  {\bibfnamefont {A.~K.}\ \bibnamefont {McMahan}}, \bibinfo {author}
  {\bibfnamefont {W.~E.}\ \bibnamefont {Pickett}}, \bibinfo {author}
  {\bibfnamefont {R.~T.}\ \bibnamefont {Scalettar}}, \bibinfo {author}
  {\bibfnamefont {M.~Y.}\ \bibnamefont {Hu}}, \ and\ \bibinfo {author}
  {\bibfnamefont {P.}~\bibnamefont {Chow}},\ }\href {\doibase
  10.1103/PhysRevLett.95.165503} {\bibfield  {journal} {\bibinfo  {journal}
  {Phys. Rev. Lett.}\ }\textbf {\bibinfo {volume} {95}},\ \bibinfo {pages}
  {165503} (\bibinfo {year} {2005})}\BibitemShut {NoStop}%
\bibitem [{\citenamefont {Carbotte}(1990)}]{Carbotte_RMP1990}%
  \BibitemOpen
  \bibfield  {author} {\bibinfo {author} {\bibfnamefont {J.~P.}\ \bibnamefont
  {Carbotte}},\ }\href {\doibase 10.1103/RevModPhys.62.1027} {\bibfield
  {journal} {\bibinfo  {journal} {Rev. Mod. Phys.}\ }\textbf {\bibinfo {volume}
  {62}},\ \bibinfo {pages} {1027} (\bibinfo {year} {1990})}\BibitemShut
  {NoStop}%
\bibitem [{\citenamefont {Allen}\ and\ \citenamefont {Mitrovi{\'
  c}}(1983)}]{AllenMitrovic1983}%
  \BibitemOpen
  \bibfield  {author} {\bibinfo {author} {\bibfnamefont {P.~B.}\ \bibnamefont
  {Allen}}\ and\ \bibinfo {author} {\bibfnamefont {B.}~\bibnamefont {Mitrovi{\'
  c}}},\ }\href {\doibase http://dx.doi.org/10.1016/S0081-1947(08)60665-7}
  {\emph {\bibinfo {title} {Theory of Superconducting Tc}}},\ \bibinfo {series}
  {Solid State Physics}, Vol.~\bibinfo {volume} {37}\ (\bibinfo  {publisher}
  {Academic Press},\ \bibinfo {year} {1983})\ pp.\ \bibinfo {pages} {1 --
  92}\BibitemShut {NoStop}%
\bibitem [{\citenamefont {Li}\ \emph {et~al.}(2016)\citenamefont {Li},
  \citenamefont {Wang}, \citenamefont {Liu}, \citenamefont {Zhang},
  \citenamefont {Hao}, \citenamefont {Pickard}, \citenamefont {Nelson},
  \citenamefont {Needs}, \citenamefont {Li}, \citenamefont {Huang},
  \citenamefont {Errea}, \citenamefont {Calandra}, \citenamefont {Mauri},\ and\
  \citenamefont {Ma}}]{Li_dissociationHS_PRBR_2016}%
  \BibitemOpen
  \bibfield  {author} {\bibinfo {author} {\bibfnamefont {Y.}~\bibnamefont
  {Li}}, \bibinfo {author} {\bibfnamefont {L.}~\bibnamefont {Wang}}, \bibinfo
  {author} {\bibfnamefont {H.}~\bibnamefont {Liu}}, \bibinfo {author}
  {\bibfnamefont {Y.}~\bibnamefont {Zhang}}, \bibinfo {author} {\bibfnamefont
  {J.}~\bibnamefont {Hao}}, \bibinfo {author} {\bibfnamefont {C.~J.}\
  \bibnamefont {Pickard}}, \bibinfo {author} {\bibfnamefont {J.~R.}\
  \bibnamefont {Nelson}}, \bibinfo {author} {\bibfnamefont {R.~J.}\
  \bibnamefont {Needs}}, \bibinfo {author} {\bibfnamefont {W.}~\bibnamefont
  {Li}}, \bibinfo {author} {\bibfnamefont {Y.}~\bibnamefont {Huang}}, \bibinfo
  {author} {\bibfnamefont {I.}~\bibnamefont {Errea}}, \bibinfo {author}
  {\bibfnamefont {M.}~\bibnamefont {Calandra}}, \bibinfo {author}
  {\bibfnamefont {F.}~\bibnamefont {Mauri}}, \ and\ \bibinfo {author}
  {\bibfnamefont {Y.}~\bibnamefont {Ma}},\ }\href {\doibase
  10.1103/PhysRevB.93.020103} {\bibfield  {journal} {\bibinfo  {journal} {Phys.
  Rev. B}\ }\textbf {\bibinfo {volume} {93}},\ \bibinfo {pages} {020103}
  (\bibinfo {year} {2016})}\BibitemShut {NoStop}%
\bibitem [{\citenamefont {Akashi}\ \emph {et~al.}(2016)\citenamefont {Akashi},
  \citenamefont {Sano}, \citenamefont {Arita},\ and\ \citenamefont
  {Tsuneyuki}}]{akashi_h3Smagneli_condmat2015}%
  \BibitemOpen
  \bibfield  {author} {\bibinfo {author} {\bibfnamefont {R.}~\bibnamefont
  {Akashi}}, \bibinfo {author} {\bibfnamefont {W.}~\bibnamefont {Sano}},
  \bibinfo {author} {\bibfnamefont {R.}~\bibnamefont {Arita}}, \ and\ \bibinfo
  {author} {\bibfnamefont {S.}~\bibnamefont {Tsuneyuki}},\ }\href@noop {}
  {\bibfield  {journal} {\bibinfo  {journal} {ArXiv e-prints}\ } (\bibinfo
  {year} {2016})},\ \Eprint {http://arxiv.org/abs/1512.06680} {arXiv:1512.06680
  [cond-mat.supr-con]} \BibitemShut {NoStop}%
\bibitem [{\citenamefont {Boeri}\ \emph {et~al.}(2004)\citenamefont {Boeri},
  \citenamefont {Kortus},\ and\ \citenamefont
  {Andersen}}]{boeri_diamond_PRL2004}%
  \BibitemOpen
  \bibfield  {author} {\bibinfo {author} {\bibfnamefont {L.}~\bibnamefont
  {Boeri}}, \bibinfo {author} {\bibfnamefont {J.}~\bibnamefont {Kortus}}, \
  and\ \bibinfo {author} {\bibfnamefont {O.~K.}\ \bibnamefont {Andersen}},\
  }\href {\doibase 10.1103/PhysRevLett.93.237002} {\bibfield  {journal}
  {\bibinfo  {journal} {Phys. Rev. Lett.}\ }\textbf {\bibinfo {volume} {93}},\
  \bibinfo {pages} {237002} (\bibinfo {year} {2004})}\BibitemShut {NoStop}%
\bibitem [{\citenamefont {Oganov}\ and\ \citenamefont
  {Glass}(2006{\natexlab{b}})}]{oganov2006crystal}%
  \BibitemOpen
  \bibfield  {author} {\bibinfo {author} {\bibfnamefont {A.~R.}\ \bibnamefont
  {Oganov}}\ and\ \bibinfo {author} {\bibfnamefont {C.~W.}\ \bibnamefont
  {Glass}},\ }\href
  {http://scitation.aip.org/content/aip/journal/jcp/124/24/10.1063/1.2210932}
  {\bibfield  {journal} {\bibinfo  {journal} {The Journal of chemical physics}\
  }\textbf {\bibinfo {volume} {124}},\ \bibinfo {pages} {244704} (\bibinfo
  {year} {2006}{\natexlab{b}})}\BibitemShut {NoStop}%
\bibitem [{\citenamefont {Lyakhov}\ \emph {et~al.}(2013)\citenamefont
  {Lyakhov}, \citenamefont {Oganov}, \citenamefont {Stokes},\ and\
  \citenamefont {Zhu}}]{lyakhov2013new}%
  \BibitemOpen
  \bibfield  {author} {\bibinfo {author} {\bibfnamefont {A.~O.}\ \bibnamefont
  {Lyakhov}}, \bibinfo {author} {\bibfnamefont {A.~R.}\ \bibnamefont {Oganov}},
  \bibinfo {author} {\bibfnamefont {H.~T.}\ \bibnamefont {Stokes}}, \ and\
  \bibinfo {author} {\bibfnamefont {Q.}~\bibnamefont {Zhu}},\ }\href
  {http://www.sciencedirect.com/science/article/pii/S0010465512004055}
  {\bibfield  {journal} {\bibinfo  {journal} {Computer Physics Communications}\
  }\textbf {\bibinfo {volume} {184}},\ \bibinfo {pages} {1172} (\bibinfo {year}
  {2013})}\BibitemShut {NoStop}%
\bibitem [{\citenamefont {Oganov}\ \emph {et~al.}(2011)\citenamefont {Oganov},
  \citenamefont {Lyakhov},\ and\ \citenamefont
  {Valle}}]{oganov2011evolutionary}%
  \BibitemOpen
  \bibfield  {author} {\bibinfo {author} {\bibfnamefont {A.~R.}\ \bibnamefont
  {Oganov}}, \bibinfo {author} {\bibfnamefont {A.~O.}\ \bibnamefont {Lyakhov}},
  \ and\ \bibinfo {author} {\bibfnamefont {M.}~\bibnamefont {Valle}},\ }\href
  {http://pubs.acs.org/doi/abs/10.1021/ar1001318} {\bibfield  {journal}
  {\bibinfo  {journal} {Accounts of chemical research}\ }\textbf {\bibinfo
  {volume} {44}},\ \bibinfo {pages} {227} (\bibinfo {year} {2011})}\BibitemShut
  {NoStop}%
\bibitem [{\citenamefont {Kresse}\ and\ \citenamefont
  {Hafner}(1993)}]{kresse1993ab}%
  \BibitemOpen
  \bibfield  {author} {\bibinfo {author} {\bibfnamefont {G.}~\bibnamefont
  {Kresse}}\ and\ \bibinfo {author} {\bibfnamefont {J.}~\bibnamefont
  {Hafner}},\ }\href
  {http://journals.aps.org/prb/abstract/10.1103/PhysRevB.47.558} {\bibfield
  {journal} {\bibinfo  {journal} {Physical Review B}\ }\textbf {\bibinfo
  {volume} {47}},\ \bibinfo {pages} {558} (\bibinfo {year} {1993})}\BibitemShut
  {NoStop}%
\bibitem [{\citenamefont {Kresse}\ and\ \citenamefont
  {Furthm{\"u}ller}(1996)}]{kresse1996efficiency}%
  \BibitemOpen
  \bibfield  {author} {\bibinfo {author} {\bibfnamefont {G.}~\bibnamefont
  {Kresse}}\ and\ \bibinfo {author} {\bibfnamefont {J.}~\bibnamefont
  {Furthm{\"u}ller}},\ }\href
  {http://www.sciencedirect.com/science/article/pii/0927025696000080}
  {\bibfield  {journal} {\bibinfo  {journal} {Computational Materials Science}\
  }\textbf {\bibinfo {volume} {6}},\ \bibinfo {pages} {15} (\bibinfo {year}
  {1996})}\BibitemShut {NoStop}%
\bibitem [{\citenamefont {Perdew}\ \emph {et~al.}(1996)\citenamefont {Perdew},
  \citenamefont {Burke},\ and\ \citenamefont
  {Ernzerhof}}]{perdew1996generalized}%
  \BibitemOpen
  \bibfield  {author} {\bibinfo {author} {\bibfnamefont {J.~P.}\ \bibnamefont
  {Perdew}}, \bibinfo {author} {\bibfnamefont {K.}~\bibnamefont {Burke}}, \
  and\ \bibinfo {author} {\bibfnamefont {M.}~\bibnamefont {Ernzerhof}},\ }\href
  {http://journals.aps.org/prl/abstract/10.1103/PhysRevLett.77.3865} {\bibfield
   {journal} {\bibinfo  {journal} {Physical review letters}\ }\textbf {\bibinfo
  {volume} {77}},\ \bibinfo {pages} {3865} (\bibinfo {year}
  {1996})}\BibitemShut {NoStop}%
\bibitem [{\citenamefont {Kresse}\ and\ \citenamefont
  {Joubert}(1999)}]{kresse1999ultrasoft}%
  \BibitemOpen
  \bibfield  {author} {\bibinfo {author} {\bibfnamefont {G.}~\bibnamefont
  {Kresse}}\ and\ \bibinfo {author} {\bibfnamefont {D.}~\bibnamefont
  {Joubert}},\ }\href
  {http://journals.aps.org/prb/abstract/10.1103/PhysRevB.59.1758} {\bibfield
  {journal} {\bibinfo  {journal} {Physical Review B}\ }\textbf {\bibinfo
  {volume} {59}},\ \bibinfo {pages} {1758} (\bibinfo {year}
  {1999})}\BibitemShut {NoStop}%
\bibitem [{\citenamefont {Bl{\"o}chl}(1994)}]{blochl1994projector}%
  \BibitemOpen
  \bibfield  {author} {\bibinfo {author} {\bibfnamefont {P.~E.}\ \bibnamefont
  {Bl{\"o}chl}},\ }\href
  {http://journals.aps.org/prb/abstract/10.1103/PhysRevB.50.17953} {\bibfield
  {journal} {\bibinfo  {journal} {Physical Review B}\ }\textbf {\bibinfo
  {volume} {50}},\ \bibinfo {pages} {17953} (\bibinfo {year}
  {1994})}\BibitemShut {NoStop}%
\end{thebibliography}%
\end{document}